 \documentclass[aps,prb,preprint,groupedaddress,amsmath,amssymb]{revtex4}
\usepackage{graphicx}
\usepackage{amscd}
\usepackage{dcolumn}
\usepackage{bm}

\newcommand{\inp}{\mathrm{in}}
\newcommand{\out}{\mathrm{out}}
\newcommand{\vf}{\bm f}

\newcommand{\ve}{\bm e}
\newcommand{\vu}{\bm u}
\newcommand{\di}{\mathrm{d}}
\newcommand{\be}{\bm e}
\newcommand{\bff}{\mathbf{f}}
\newcommand{\brr}{\mathbf{r}}
\newcommand{\bq}{\mathbf{q}}
\newcommand{\bk}{\mathbf{k}}
\newcommand{\bp}{\mathbf{p}}
\newcommand{\bx}{\mathbf{x}}

\newcommand{\bu}{\mathbf{u}}
\newcommand{\ba}{\mathbf{a}}
\newcommand{\bb}{\mathbf{b}}
\newcommand{\bee}{\mathbf{e}}
\newcommand{\bvarphi}{\bm \varphi}
\newcommand{\brho}{\bm \rho}
\newcommand{\vkappa}{\bm \kappa}
\newcommand{\vPsi}{\bm \Psi}

\begin{document}
\title{The reflection of a Maxwell-Gaussian beam by a planar surface}
\author{A. Aiello}
\author{J. P. Woerdman}
\affiliation{Huygens Laboratory, Leiden University\\
P.O.\ Box 9504, 2300 RA Leiden, The Netherlands}
\begin{abstract}
The reflection of a three-dimensional vectorial Maxwell-Gaussian beam by a planar surface is studied. The surface is characterized by its complex reflection coefficients $r_s(\bk)$ and $r_p(\bk)$ for TE and TM electromagnetic plane waves of wavevector $\bk$, respectively. The field impinging upon the reflecting surface is modeled as a quasi-monochromatic fundamental Gaussian beam suitably modified in order to satisfy Maxwell equations (Maxwell-Gaussian beam). Analytical expressions, correct up to the second order in a perturbation expansion, are given for the reflected electric and magnetic field, respectively. We found that first order terms in the perturbation expansion account for a longitudinal shift (Goos-H\"{a}nchen effect) of the whole reflected beam, while second order terms modifies the transverse shape of the beam which is, at this order, no longer cylindrically symmetric.
\end{abstract}
\maketitle
\thispagestyle{empty}
\section{Introduction}
Plane waves of the form $\vu \exp( i \bk \cdot \brr - i\omega t)$, (with $ \vu \cdot \bk =0$ and $\omega = c |\bk |$), are  solutions of the Maxwell equations that show a serious problem: They are physically impossible since posses an infinite amount of energy. However, they are very easy to handle and, thus, widely used in the physics community. Moreover, an electromagnetic field of a given arbitrary shape, can always be written as a \emph{linear} superposition of plane waves, hence, for example, it is enough to know how a plane wave propagate across some medium, to know how the whole field does.  For this reasons, even in advanced textbooks \cite{BandWBook}, the reflection coefficients associated to the interface between two media, are calculated in terms of the amplitudes of incident and reflected \emph{plane waves}. However, since in our real world plane waves of infinite transverse extension  do not exist, but only finite-transverse-size light beams,   non-specular reflection effects are expected \cite{KandS,Tamir,Nasalski,LandM,Barton,Gragg} and, actually, occur. The most known effects are the  Goos-H\"{a}nchen \cite{CandQ,McGandC,LaiEtAl,LeungEtAl} and the Imbert-Fedorov \cite{CandI,PillonEtAl} longitudinal and transverse shifts \cite{Li}, respectively.

In these Notes we study in a detailed and didactic manner the non-specular effects occurring when a Maxwell-Gaussian beam \cite{EandS} impinges upon an arbitrary planar surface  characterized by its complex reflection coefficients \cite{EsquivelEtAl}.
\section{Geometric reflection}
Before discussing the \emph{physical} process of reflection of light
by a planar surface, let us consider some general characteristic of
\emph{geometric} reflection by an ideal planar mirror (for a short and simple introduction to mirror symmetry applied to electromagnetism see, e.g., ref. \cite{Yao}).  Let $K =
(Oxyz)$ be a Cartesian reference frame whose axes $x,y,z$ are
specified by the three unit basis vectors $\{\be_x, \be_y,\be_z \}$,
respectively, and let $z = 0$ be the equation of the reflecting
planar surface. The geometry of the problem at hand is illustrated
in fig. 1.  We define geometric reflection (or mirror symmetry with respect to the plane $z=0$) in an operative
fashion, as follows. If a point $P$ has coordinates $\brr = \be_x x
+ \be_y y + \be_z z$ in $K$, then the mirror image
point $\widetilde{P}$ has coordinates $\widetilde{\brr} = \be_x x +
\be_y y - \be_z z$.
Thus, the mirror image of a \emph{scalar} field $\phi(x,y,z)$,
is simply a new field $\widetilde{\phi}(x,y,z)$ defined as
\begin{equation}\label{gh10}
\phi(x,y,z) \rightarrow \widetilde{\phi}(x,y,z) = \phi(x,y,-z).
\end{equation}
\begin{flushright}
\begin{figure}[]
\includegraphics[angle=0,width=7truecm]{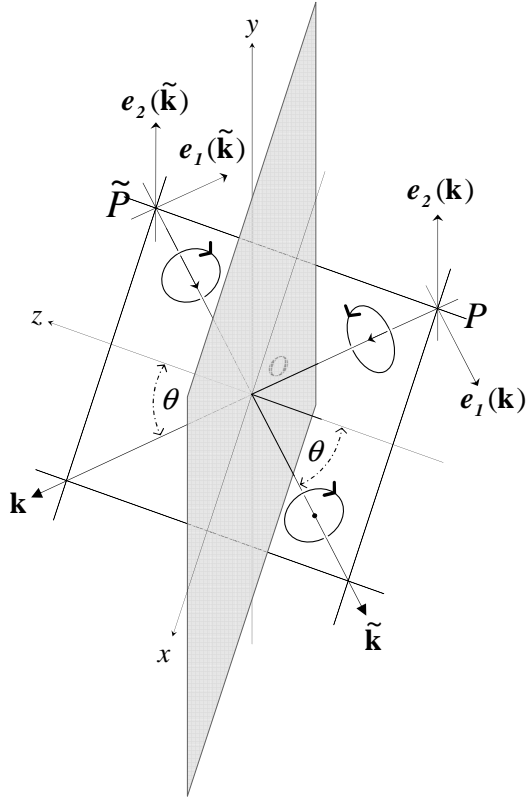}
\caption{\label{fig:1} Illustrating the geometric reflection (or mirror symmetry) of a
right-handed circularly (RHC) polarized plane wave by an ideal
mirror. The grey plane of equation $z=0$ is the ideal mirror
surface. }
\end{figure}
\end{flushright}
Next, we must consider geometric reflection of \emph{vector}
fields. To this end we study, without loss of generality, the
reflection of a circularly polarized plane wave with wave vector
$\bk$, impinging upon the $xy$-plane from $z<0$, as shown in Fig. 1. Geometric
reflection changes $\bk$ in $\widetilde{\bk}$ and transforms the image of a
right hand in the image of a left hand. Thus,  the mirror image of a
\emph{right-hand} circularly polarized plane wave, must be a
\emph{left-hand} circularly polarized plane wave, and vice versa.
These statements can be straightforwardly transformed in formulas as follows:
Let us write the incident plane wave of unit amplitude as
\begin{equation}\label{gh20}
\mathbf{A}^\mathrm{inc}(\brr,t) =   \frac{\ve_1(\bk) + i
\ve_2(\bk) }{2^{1/2}} \exp \bigr[i \bk \cdot \brr - i \omega(\bk)
t \bigl],
\end{equation}
where $\omega(\bk) = c |\bk|$ and the orthogonal basis vectors $\{
\ve_i(\bk) \}_{i=1}^3$ are defined as in Fig. 1 and Appendix A. According to the
definition of geometric reflection  given above,
the reflected field $\mathbf{A}^\mathrm{ref}(\brr,t)$ is still a
plane wave of unit amplitude with equation
\begin{equation}\label{gh30}
\mathbf{A}^\mathrm{ref}(\brr,t)  =   \frac{\ve_1(\widetilde{\bk})
- i \ve_2(\widetilde{\bk}) }{2^{1/2}} \exp \bigl[i \widetilde{\bk}
\cdot \brr - i \omega(\bk) t \bigr],
\end{equation}
where  $\widetilde{\bk} = \be_x k_{x} + \be_y k_{y} - \be_z k_{z}$
lies on the plane of incidence defined as the common plane of
${\bk} = \be_x k_{x} + \be_y k_{y} + \be_z k_{z}$ and $\ve_z$.
Note that  $\widetilde{\bk} \cdot \widetilde{\bk} = \bk \cdot
\bk$, therefore  $|\widetilde{\bk}| = |\bk| \Rightarrow
\omega(\widetilde{\bk}) = \omega(\bk)$.

It is easy to generalize the results above to the case of an
arbitrary incident field. By definition, such a field can always be written
as a plane waves expansion of the form:
\begin{eqnarray}\label{gh40}
\mathbf{A}^\mathrm{inc}(\brr,t) =  \int \left[ \ve_1(\bk)
a_{1}(\bk)+ \ve_2(\bk) a_{2}(\bk) \right] \exp \left[i {\bk} \cdot
\brr - i \omega(\bk) t \right] \di^3 k.
\end{eqnarray}
Since from Eqs. (\ref{gh20})-(\ref{gh30}) we know how each plane
wave making $\mathbf{A}^\mathrm{inc}(\brr,t)$ transforms under
geometric reflection, we can write at once
\begin{eqnarray}\nonumber
\mathbf{A}^\mathrm{ref}(\brr,t) & = & \int \bigl[
\ve_1(\widetilde{\bk}) a_{1}(\bk) -\ve_2(\widetilde{\bk})
a_{2}(\bk)  \bigr] \exp \bigl[i
{\widetilde{\bk}} \cdot \brr - i \omega(\bk) t \bigr] \di^3 k \\
\label{gh50}
& = & \sum_{\lambda=1}^2\int \ve_\lambda(\widetilde{\bk}) \,
r_\lambda \, a_{\lambda}(\bk) \exp \bigl[i {\bk \cdot  \widetilde{\brr}}
- i \omega(\bk) t \bigr] \di^3 k,
\end{eqnarray}
where we have defined the reflection coefficients
\begin{eqnarray}\label{gh60}
r_\lambda =  \left\{%
\begin{array}{ll}
    + 1, & \lambda = 1, \\
    -1, & \lambda =2, \\
\end{array}%
\right.
\end{eqnarray}
and we have used the property $\widetilde{\bk} \cdot \brr = \bk \cdot  \widetilde{\brr}$.
Note that according to Eqs. (\ref{gh20})-(\ref{gh30}), the
amplitudes $a_\lambda(\bk)$ do not change by geometric reflection, that is $
a_\lambda(\bk) \; { \backslash \! \! \! \! \! \! \rightarrow} \;
a_\lambda(\widetilde{\bk})$.
\section{Physical reflection}
Now, let us consider the case of reflection by an actual
physical planar surface characterized by the reflection amplitude
coefficients (with respect to the vacuum) $r_s({\bk}), \,
r_p({\bk})$ for $s$-polarized waves (or $\mathrm{TE}$, that is plane waves with the electric field
\emph{orthogonal} to the plane of incidence), and $p$-polarized
waves (or $\mathrm{TM}$, that is plane waves with the electric field \emph{parallel} to the plane of
incidence), respectively. From our choice (see appendix A) for the basis vectors
$\{ \ve_i(\bk) \}_{i=1}^3$, with $\ve_1(\bk)$ and $\ve_2(\bk)$  parallel and orthogonal with respect to the plane of incidence, respectively,  it follows that
$r_p(\bk) \equiv r_1(\bk)$, $r_s(\bk) \equiv r_2(\bk)$. For
an  homogeneous medium with complex-valued dielectric constant $\hat{\varepsilon}$
and refractive index\footnotemark[1]{} $\hat{n} = \sqrt{\hat{\varepsilon}}\equiv n + i \kappa$, $(n,\kappa \in \mathbb{R}) $\footnotetext[1]{Warning: Note that the present definition of $\hat{n}$ is different from the one given in Chap. XIII of the book by Born and Wolf, where the authors write $\hat{n} = n(1 + i \kappa)$} , we have
\begin{align}\label{gh70}
 r_1(\bk) &=  \frac{\hat{\varepsilon}k_z - k_{m z}}{\hat{\varepsilon}k_z + k_{m z}}, \\
\label{gh145}  r_2(\bk) &= \frac{k_z - k_{m z}}{k_z + k_{m z}},
\end{align}
where $\bk = \ve_x k_x + \ve_y k_y + \ve_z k_{z}$, and $\bk_m = \ve_x k_x + \ve_y k_y + \ve_z k_{m z}$ is the wave vector of the plane wave transmitted into the medium, as given by the Snell law, and
\begin{eqnarray}
 \label{gh147}  k_{m z} = \sqrt{\hat{\varepsilon} k_z^2 + (\hat{\varepsilon} -1)(k_x^2 +k_y^2)},
\end{eqnarray}
is the (generally complex-valued) $z$-component of $\bk_m$.

 Once we know the reflection coefficients
$\{r_\lambda(\bk)\}_{\lambda=1}^2$ associated to a single plane
wave with wave vector $\bk$, we can easily determine the behavior
under reflection of an arbitrary field just by letting $r_\lambda
\rightarrow r_\lambda(\bk)$ in Eq. (\ref{gh50}):
\begin{eqnarray}\label{gh80}
\mathbf{A}^\mathrm{ref}(\brr,t) = \sum_{\lambda=1}^2\int
\ve_\lambda(\widetilde{\bk}) r_\lambda(\bk) a_{\lambda}(\bk) \exp
\bigl[i {\widetilde{\bk}} \cdot \brr - i \omega(\bk) t \bigr]
\di^3 k,
\end{eqnarray}
Equation (\ref{gh80}) is perfectly general and, therefore, of
limited usefulness. However, much additional work can be done if
we consider the \emph{actual} experimental situation where the
incident field is a quasi-monochromatic narrow beam  directed
along $\bk_0$ with central frequency $\omega_0 = c |\bk_0|$. Such
a beam can be represented by an envelope vector field ${\bm
\Psi}^\mathrm{inc}(\brr,t)$ modulating a carrier plane wave with
wave vector $\bk_0$ and frequency $\omega_0$:
\begin{eqnarray}\label{gh90}
\mathbf{A}^\mathrm{inc}(\brr,t) = {\bm \Psi}^\mathrm{inc}(\brr,t)
\exp \left[i ( {{\bk_0}} \cdot \brr - \omega_0 t ) \right].
\end{eqnarray}
The envelope ${\bm \Psi}^\mathrm{inc}(\brr,t)$ is easily
determined by rewriting Eq. (\ref{gh90}) as
\begin{eqnarray}\label{gh100} \nonumber
\mathbf{A}^\mathrm{inc}(\brr,t) & = & \exp \left[i ( \bk_0\cdot
\brr - \omega_0 t ) \right] \\ \nonumber
& & \times  \sum_{\lambda =1}^2 \int \, \Bigl\{ \Bigr.
\be_{\lambda}(\bk) a_{\lambda}(\bk) \exp \left[i \bigr(
{\bk} - \bk_0 \bigl) \cdot \brr  \right] \\
& & \Bigr. \times \exp \left[ - i \bigl( \omega(\bk) - \omega_0
\bigr) t \right] \Bigl\} \mathrm{d}^3 k .
\end{eqnarray}
The very same procedure can be executed for the reflected beam
obtaining an envelope vector field ${\bm
\Psi}^\mathrm{ref}(\brr,t)$ modulating a carrier plane wave with
wave vector $\widetilde{\bk}_0$ and frequency $\omega_0$:
\begin{eqnarray}\label{gh110}
\mathbf{A}^\mathrm{ref}(\brr,t) = {\bm \Psi}^\mathrm{ref}(\brr,t)
\exp \left[i ( \widetilde{\bk}_0 \cdot \brr - \omega_0 t )
\right],
\end{eqnarray}
where
\begin{eqnarray}\nonumber
{\bm
\Psi}^\mathrm{ref}(\brr,t) & = &   \sum_{\lambda =1}^2 \int \, \Bigl\{ \Bigr.
\be_{\lambda}(\widetilde{\bk}) r_{\lambda}(\bk) a_{\lambda}(\bk)
\exp \left[i \bigr( \widetilde{\bk} - \widetilde{\bk}_0 \bigl)
\cdot \brr \right] \\ \label{gh120}
& & \Bigr. \times \exp \left[ - i \bigl( \omega(\bk) - \omega_0
\bigr) t \right] \Bigl\} \mathrm{d}^3 k .
\end{eqnarray}
Since $\bigr( \widetilde{\bk} - \widetilde{\bk}_0 \bigl) \cdot
\brr = \bigr( \bk - \bk_0 \bigl) \cdot \widetilde{\brr}$ we can
write
\begin{eqnarray}\nonumber
{\bm \Psi}^\mathrm{ref}(\brr,t) &= & \sum_{\lambda =1}^2 \int \,
\Bigl\{ \Bigr. \be_{\lambda}(\widetilde{\bk}) r_{\lambda}(\bk)
a_{\lambda}(\bk) \exp \left[i \bigr(
\bk - \bk_0 \bigl) \cdot \widetilde{\brr}  \right] \\
\nonumber
& & \Bigr. \times \exp \left[ - i \bigl( \omega(\bk) - \omega_0
\bigr) t \right] \Bigl\} \mathrm{d}^3 k \\ \nonumber
&= & \sum_{\lambda =1}^2 \int \, \Bigl\{ \Bigr.
\be_{\lambda}(\widetilde{\bk}_0 + \widetilde{\bq})
r_{\lambda}(\bk_0 + \bq) a_{\lambda}(\bk_0
+ \bq) \exp \left(i \bq  \cdot \widetilde{\brr}  \right) \\
\label{gh130}
& & \Bigr. \times \exp \left[ - i \delta \omega(\bk_0, \bq) t
\right] \Bigl\} \mathrm{d}^3 q,
\end{eqnarray}
were in the second line we have  changed the variables of
integration according to:
\begin{eqnarray}\label{gh140}
\bk \rightarrow \bq = \bk - \bk_0, \qquad \di^3 k = \di^3q,
\end{eqnarray}
and we have defined
\begin{eqnarray}\label{gh142}
\delta \omega(\bk_0, \bq) \equiv \omega(\bk_0 + \bq) - \omega_0 .
\end{eqnarray}
\begin{flushright}
\begin{figure}[h]
\includegraphics[angle=0,width=5truecm]{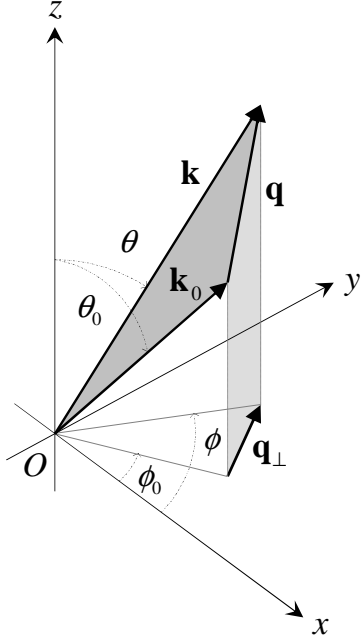}
\caption{\label{fig:2} Illustrating the geometric meaning of the
change of integration variables $\bk \rightarrow \bq = \bk - \bk_0$.
}
\end{figure}
\end{flushright}
By hypothesis, we are considering a narrow beam, therefore we
expect that the difference vector $\bq = \bk - \bk_0$ would be
small as compared to $\bk_0$: $|\bq| / k_0 \ll 1$ \cite{DandG}.
Then, we can Taylor expand the reflection coefficients
$r_{\lambda}(\bk_0 + \bq)$ around $\bq = {\bm 0}$. To this end,
first we rewrite $r_{\lambda}(\bk_0 + \bq)$ as
\begin{eqnarray}\nonumber
r_{\lambda}(\bk_0 + \bq) & = & \left| r_{\lambda}(\bk_0 + \bq)
\right| \exp \left[ i \phi_\lambda (\bk_0 + \bq)\right] \\
\label{gh150}
& \equiv & R_{\lambda}(\bk_0 + \bq) \exp \left[ i \phi_\lambda
(\bk_0 + \bq)\right],
\end{eqnarray}
and then we make a  Taylor expansion for both the amplitudes
$R_{\lambda}(\bk_0 + \bq)$ and the phases $  \phi_\lambda (\bk_0 +
\bq)$, separately.
For the latter we have, up to  second order terms \cite{Vandegrift}:
\begin{eqnarray}\label{gh220} \nonumber
\phi_\lambda (\bk_0 + \bq) & = & \phi_\lambda (\bk_0) + \left.
\frac{\partial \phi_\lambda}{ \partial k_i} \right|_{\bk = \bk_0}
q_i + \frac{1}{2} \left. \frac{\partial^2 \phi_\lambda}{ \partial
k_i
\partial k_j} \right|_{\bk = \bk_0} q_i q_j + \ldots
\\
& \equiv & \phi_\lambda (\bk_0) + \bvarphi_\lambda \cdot \bq +
\frac{1}{2} \bq \cdot \mathcal{F}_\lambda \bq  + \ldots,
\end{eqnarray}
where summation over repeated indices is understood. The
``displacement'' vectors $\bvarphi_\lambda =\bvarphi_\lambda(\bk_0)$
and the symmetric $3 \times 3$ tensors $\mathcal{F}_\lambda =
\mathcal{F}_\lambda(\bk_0)$ are defined as
%
\begin{align}\label{gh230}
\varphi^i_\lambda (\bk_0)&= \left[
\frac{\partial}{ \partial k_i}  \phi_\lambda(\bk) \right]_{\bk = \bk_0}, \\
\mathcal{F}_\lambda^{ij} (\bk_0)&= \left[ \frac{\partial^2 }{
\partial k_i
\partial k_j} \phi_\lambda(\bk) \right]_{\bk = \bk_0},
\end{align}
%
$(i,j \in \{
1,2,3\})$, respectively. A similar Taylor expansion can be also performed for
the amplitudes $R_{\lambda}(\bk_0 + \bq)$ after their
exponentiation:
\begin{eqnarray}\label{gh240}
R_{\lambda}(\bk_0 + \bq) = \exp \left[ \ln R_{\lambda}(\bk_0 +
\bq) \right],
\end{eqnarray}
where
\begin{eqnarray}\label{gh250} \nonumber
\ln R_{\lambda}(\bk_0 + \bq) & = & \ln R_{\lambda}(\bk_0) + q_i
\left[\frac{1}{R_\lambda(\bk)} \frac{\partial  R_\lambda}{
\partial k_i} \right]_{\bk = \bk_0}
\\
 &  &  + \frac{1}{2}  q_i q_j
\left[ \frac{1}{R_\lambda(\bk)} \frac{\partial^2 R_\lambda }{
\partial k_i \partial k_j} -  \frac{1}{R_\lambda^2(\bk)} \frac{\partial R_\lambda }{
\partial k_i} \frac{\partial R_\lambda }{
\partial k_j} \right]_{\bk = \bk_0} + \ldots
\\
& \equiv &  \ln R_{\lambda}(\bk_0) + \brho_\lambda \cdot \bq +
\frac{1}{2} \bq \cdot \mathcal{R}_\lambda \bq  + \ldots.
\end{eqnarray}
The ``amplitude modulation'' vectors $\brho_\lambda
=\brho_\lambda(\bk_0)$ and the symmetric $3 \times 3$ tensors
$\mathcal{R}_\lambda = \mathcal{R}_\lambda(\bk_0)$ are defined as
\begin{eqnarray}\label{gh260}
\rho^i_\lambda (\bk_0)&= &\frac{1}{R_\lambda(\bk_0)}\left[
\frac{\partial }{
\partial k_i}R_\lambda(\bk) \right]_{\bk = \bk_0}, \\
\mathcal{R}_\lambda^{ij} (\bk_0)&=& \frac{1}{R_\lambda(\bk_0)}
\left[ \frac{\partial^2  }{
\partial k_i \partial k_j}R_\lambda(\bk) \right]_{\bk = \bk_0} - \rho^i_\lambda
(\bk_0) \rho^j_\lambda (\bk_0),
%
%
\end{eqnarray}
respectively. It is important to stress that all partial
derivatives in Eqs. (\ref{gh220})-(\ref{gh260}) are understood to
be \emph{whole-partial} derivatives \cite{Brownstein}, that is, if $f = f(\bk,
\omega(\bk))$ is a smooth function that depends both explicitly
upon $\bk$ via $f(\bk, \omega(\bk))$, and implicitly upon $\bk$
via $\omega(\bk)$, then with $\partial / \partial k_i$ we mean
\begin{eqnarray}\label{gh270}
\frac{\partial}{\partial k_i} f(\bk, \omega(\bk)) \Leftrightarrow
\left[ \frac{\partial}{\partial k_i} + \frac{\partial
\omega}{\partial k_i} \frac{\partial}{\partial \omega}\right]
f(\bk, \omega(\bk)).
\end{eqnarray}
Now, by combining Eq. (\ref{gh150}) with Eq. (\ref{gh220}) and Eq.
(\ref{gh250}), we obtain
\begin{eqnarray}\nonumber
r_{\lambda}(\bk_0 + \bq) & = & R_{\lambda}(\bk_0) \exp \left(
\brho_\lambda \cdot \bq +
\frac{1}{2} \bq \cdot \mathcal{R}_\lambda \bq  + \ldots\right)\\
\nonumber
&  & \times \exp  \left( i \phi_\lambda (\bk_0) + i \bvarphi_\lambda
\cdot \bq + \frac{i}{2} \bq \cdot \mathcal{F}_\lambda \bq  +
\ldots \right),
\\
\label{gh280}
& = & r_{\lambda}(\bk_0) \exp  \left[  \bigl(\brho_\lambda + i \bvarphi_\lambda
 \bigr) \cdot \bq + \frac{1}{2} \bq \cdot \bigl( \mathcal{R}_\lambda
 + i \mathcal{F}_\lambda \bigr) \bq  + \ldots
\right],
\end{eqnarray}
where $r_{\lambda}(\bk_0) = R_{\lambda}(\bk_0) \exp[i
\phi_{\lambda}(\bk_0)]$. Finally, we substitute this expression into
Eq. (\ref{gh130}) to obtain
\begin{eqnarray}\nonumber
{\bm \Psi}^\mathrm{ref}(\brr,t)
&= & \sum_{\lambda =1}^2 \int \, \left\{ \Bigr.
\be_{\lambda}(\widetilde{\bk}_0 + \widetilde{\bq})
r_{\lambda}(\bk_0 + \bq) a_{\lambda}(\bk_0
+ \bq) \exp (i \widetilde{\bq} \cdot \brr) \right. \\
\nonumber
& & \Bigr. \times \exp \left[ - i \delta \omega(\bk_0, \bq) t
\right] \Bigl\} \mathrm{d}^3 q \\ \nonumber
& = &\sum_{\lambda =1}^2  r_{\lambda}(\bk_0)\int \,
\be_{\lambda}(\widetilde{\bk}_0 + \widetilde{\bq})
 a_{\lambda}(\bk_0
+ \bq) \exp \left[ - i \delta \omega(\bk_0, \bq) t
\right] \\%
\label{gh290}
& &  \times \exp  \left\{ i\bq \cdot \left[  \bigl(\widetilde{\brr} + \bvarphi_\lambda - i
\brho_\lambda \bigr)  + \frac{1}{2} \bigl(
\mathcal{F}_\lambda - i \mathcal{R}_\lambda \bigr) \bq  + \ldots \right]
\right\} \mathrm{d}^3 q.
\end{eqnarray}
This is our final expression. One can see that  the whole beam
undergoes both a linear $\bvarphi_\lambda - i
\brho_\lambda $ and  non-linear
$(\mathcal{F}_\lambda - i \mathcal{R}_\lambda)\bq $ complex displacement.
\subsection{Reflection of vector Gaussian beams}
In this subsection, we consider an incident vector field tailored
as a Gaussian beam. From Appendix D, we know that such a  field
can be described as
\begin{eqnarray}\label{gh300}
\mathbf{E}^\mathrm{inc}(\brr,t) = \frac{1}{(2 \pi)^{3/2}}
\sum_{\lambda=1}^2 \int \ve_\lambda(\bk) a_\lambda(\bk) \exp[i
(\bk \cdot \brr - \omega t)] \di^3 k,
\end{eqnarray}
where
\begin{eqnarray}\label{gh310}
a_\lambda(\bk) = a(\bk) \ve_\lambda(\bk) \cdot \vf \equiv a(\bk)
f_\lambda(\bk).
\end{eqnarray}
The complex-valued unit vector $\vf$ specifies the orientation of
the polarizer that selects the polarization of the incident beam.
 Equation (\ref{gh300}) furnishes an expression valid for the electric field only. However, it is not difficult to see that  we can also write the magnetic field as
\begin{eqnarray}\label{gh312}
\mathbf{B}^\mathrm{inc}(\brr,t) = \frac{1}{(2 \pi)^{3/2}}
\sum_{\lambda=1}^2 \int \ve_\lambda(\bk) b_\lambda(\bk) \exp[i
(\bk \cdot \brr - \omega t)] \di^3 k,
\end{eqnarray}
were
\begin{eqnarray}\label{gh310b}
\sum_{\lambda=1}^2 \ve_\lambda(\bk) b_\lambda(\bk) = \frac{1}{\omega}  \left[ \bk \times  \sum_{\lambda=1}^2 \ve_\lambda(\bk)  a_\lambda(\bk)\right].
\end{eqnarray}
Explicitly, we have
\begin{eqnarray}\label{gh312b}
b_1(\bk) &= &- \frac{1}{c}  a( \bk ) f_2(\bk), \\ \label{gh313b}
b_2(\bk) &= & \frac{1}{c}  a( \bk ) f_1(\bk),
\end{eqnarray}
were $c$ is the speed of light in vacuum. Thus, if with $\mathbf{A}^\mathrm{inc}(\brr,t)$ we denotes either the electric or the magnetic field, we can write it as
\begin{eqnarray}\label{gh330}
\mathbf{A}^\mathrm{inc}(\brr,t) & = &  \frac{1}{(2 \pi)^{3/2}}
\sum_{\mu =1}^2 \int
 \ve_\mu(\bk)  g_\mu(\bk) \exp[i (\bk \cdot \brr - \omega t)] \di^3 k
\end{eqnarray}
with
\begin{eqnarray}\label{gh340}
g_\mu(\bk) = a(\bk) \sum_{\nu =1}^2 \ell_{\mu \nu} \ve_\nu(\bk) \cdot \vf,
\end{eqnarray}
where we have defined the $2 \times 2$ matrix $\ell = [\ell_{\mu \nu}]$ as
\begin{equation}\label{gh342}
 \ell = \left\{
 \begin{array}{cc}
 \left(
         \begin{array}{cc}
           1 & 0 \\
           0 & 1 \\
         \end{array}
       \right) \equiv \ell^E, \qquad \mathrm{for \; the \; electric \; field},
\\
\displaystyle{ \frac{1}{c} } \left(
         \begin{array}{cc}
           0 & -1 \\
           1 & 0 \\
         \end{array}
       \right) \equiv \ell^B, \qquad  \mathrm{for \; the \; magnetic \; field}.
\end{array}
       \right.
\end{equation}
Since we want to express $\mathbf{A}^\mathrm{inc}(\brr,t)$ in
terms of the Fourier transform $F(\bp)$ of the Gaussian beam
$f(\brr)$,  we follows the procedure explained in Appendix C and make the change of variables
$\bk \rightarrow \bq$, where  $\bk = \bk_0 + \bq$, thus obtaining:
\begin{eqnarray}\nonumber
\mathbf{A}^\mathrm{inc}(\brr,t) & = &  \frac{1}{(2 \pi)^{3/2}} \sum_{\mu =1}^2 \int
\ve_\mu(\bk_0 + \bq) g_\mu(\bk_0 + \bq)  \exp[i( \bk_0 + \bq)\cdot \brr - i
\omega(\bk_0 + \bq)t] \di^3 q \\
\label{gh350}
& = &  \frac{e^{i (\bk_0 \cdot \brr - \omega_0 t)}}{(2 \pi)^{3/2}} \sum_{\mu =1}^2 \int
\ve_\mu(\bk_0 + \bq) g_\mu(\bk_0 + \bq)  \exp[i\bq\cdot \brr - i
\delta \omega(\bk_0, \bq)t] \di^3 q ,
\label{gh351}
\end{eqnarray}
where $\delta \omega(\bk_0, \bq)$ is defined by Eq. (\ref{gh142}).
Under  reflection this equation changes according the
following rules: $\brr \rightarrow \widetilde{\brr}$, for the scalar
part of the field, and
\begin{equation}\label{gh351b}
\sum_{\mu,\nu =1}^2 \ve_\mu(\bk_0 + \bq) \ell_{\mu \nu} \left[ \ve_\nu(\bk) \cdot \vf \right] \rightarrow
\sum_{\mu,\nu =1}^2 \ve_\mu(\widetilde{\bk}_0 + \widetilde{\bq}) \ell_{\mu \nu} \left[ \ve_\nu(\bk) \cdot \vf \right] r_\nu(\bk_0 + \bq),
\end{equation}
%
%
 for the vector part of the field. Thus, from Eq.
(\ref{gh350}) it readily follows
\begin{eqnarray}\nonumber
\mathbf{A}^\mathrm{ref}(\brr,t) & = &  \frac{e^{i (\bk_0 \cdot \widetilde{\brr} - \omega_0 t)}}{(2 \pi)^{3/2}} \sum_{\mu =1}^2 \int
\ve_\mu(\widetilde{\bk}_0 + \widetilde{\bq})r_\nu(\bk_0 + \bq) g_\mu(\bk_0 + \bq)\\
 \nonumber
& &   \times \exp[i\bq\cdot \widetilde{\brr} - i
\delta \omega(\bk_0, \bq)t] \di^3 q\\
 \nonumber
 & =  &\Biggl\{ \frac{e^{i (\bk_0 \cdot \widetilde{\brr} - \omega_0 t)}}{(2 \pi)^{3/2}} \sum_{\mu, \nu =1}^2 \ell_{\mu \nu} \int
\ve_\mu(\widetilde{\bk}_0 + \widetilde{\bq}) \ve_\nu(\bk_0 + \bq)r_\nu(\bk_0 + \bq) \Biggr.\\
 \nonumber
& &  \quad \Biggl. \times a(\bk_0 + \bq) \exp[i\bq\cdot \widetilde{\brr} - i
\delta \omega(\bk_0, \bq)t] \di^3 q \Biggr\} \cdot \vf \\ \label{gh352}
& \equiv & \mathcal{M}^\mathrm{ref}(\brr,t) \cdot \vf,
\end{eqnarray}
where $\mathcal{M}^\mathrm{ref}(\brr,t)$ is a $3 \times 3$ space-time dependent matrix.
From Eq. (\ref{b320}) it follows that
\begin{equation*}
a(\bk_0 + \bq) = F(D \bq) \exp[-i (\bk_0 + \bq )\cdot \brr_0],
\end{equation*}
which can be used in Eq. (\ref{gh352}) to obtain
\begin{eqnarray}\nonumber
\mathcal{M}^\mathrm{ref}(\brr,t)
 & =  & \frac{e^{i [\bk_0 \cdot ( \widetilde{\brr} - \brr_0) - \omega_0 t]}}{(2 \pi)^{3/2}} \sum_{\mu, \nu =1}^2 \ell_{\mu \nu} \int
\ve_\mu(\widetilde{\bk}_0 + \widetilde{\bq}) \ve_\nu(\bk_0 + \bq)r_\nu(\bk_0 + \bq) \\
 \nonumber
& &  \quad  \times F(D \bq) \exp[i\bq\cdot ( \widetilde{\brr} - \brr_0) - i
\delta \omega(\bk_0, \bq)t] \di^3 q \\ \nonumber
 \\
 \nonumber
& =  & \frac{e^{i ( k_0 z'' - \omega_0 t)}}{(2 \pi)^{3/2}} \sum_{\mu, \nu =1}^2 \ell_{\mu \nu} \int
\ve_\mu(\widetilde{\bk}_0 + \widetilde{\bq}) \ve_\nu(\bk_0 + \bq)r_\nu(\bk_0 + \bq) \\
 \label{gh354}
& &  \quad  \times F(D \bq) \exp[i D \bq\cdot X \bx'' - i
\delta \omega(\bk_0, \bq)t] \di^3 q,
\end{eqnarray}
where we have used Eqs. (\ref{b530}-\ref{b540},\ref{b570}) to rewrite $\bk_0
\cdot (\widetilde{\brr} - \brr_0) = \widetilde{\bk}_0 \cdot (\brr -
\widetilde{ \brr}_0) = k_0  z''$, and
\begin{eqnarray}\label{gh370}
\bq \cdot(\widetilde{\brr} -\brr_0) = D \bq \cdot D(\widetilde{\brr}
-\brr_0) = D \bq \cdot   X \widetilde{D} (\brr - \widetilde{\brr}_0
) = D \bq \cdot   X \bx'' .%
\end{eqnarray}
At this point, Eq. (\ref{gh354}) is  exact but not very useful.
However, we can exploit the hypothesis that the impinging light beam
is modeled as a narrow Gaussian beam, and
make a Taylor expansion about $\bk = \bk_0$ (or, equivalently, $\bq
= {\bm 0}$),  of all relevant quantities in Eq. (\ref{gh354}). In
practice, we substitute into Eq. (\ref{gh354}),  the expansion
(\ref{gh280}):
\begin{eqnarray} \nonumber
r_{\nu}(\bk_0 + \bq) & \cong &
r_{\nu}(\bk_0) \exp  \left[ \bq \cdot \bigl(\brho_\nu + i \bvarphi_\nu  \bigr)   + \frac{1}{2} \bq \cdot \bigl(
\mathcal{R}_\nu + i \mathcal{F}_\nu \bigr) \bq
\right]\\ \nonumber
& \equiv &
r_{\nu}(\bk_0) \exp  \left( \frac{ \bq }{k_0}\cdot {\bm \upsilon}_\nu   + \frac{1}{2 k_0^2} \bq \cdot \mathcal{U}_\nu \bq
\right)\\ \label{gh385}
& = &
r_{\nu}(\bk_0) \exp  \left( \frac{q_i}{k_0} \upsilon_\nu^i + \frac{1}{2 k_0^2}q_i q_j  \mathcal{U}_\nu^{ij} ,
\right),
\end{eqnarray}
where we have chosen to truncate the Taylor
series of the amplitudes $R_{\nu}(\bk_0 + \bq)$ at the second
order terms, and we have defined $ {\bm \upsilon}_\nu \equiv k_0 (\brho_\nu + i \bvarphi_\nu)$, and $ \mathcal{U}_\nu \equiv k_0^2(\mathcal{R}_\nu + i \mathcal{F}_\nu)$. In practice,
 we have kept all the second order terms with the aim to deal
with Gaussian integrals which are analytically integrable.
Moreover, we  use the results of Appendix F to write the second-order Taylor expansions of the diadic forms
$\ve_\mu(\widetilde{\bk}_0 + \widetilde{\bq})
\ve_\nu({\bk}_0 + {\bq})$ as
\begin{eqnarray}\label{gh390}
\ve_\mu(\widetilde{\bk}_0 + \widetilde{\bq})
\ve_\nu({\bk}_0 + {\bq}) \simeq \mathbf{E}_{\mu \nu} + \frac{q_i}{k_0} \mathbf{E}^{(i)}_{\mu \nu} + \frac{1}{2 k_0^2} q_i q_j \mathbf{E}^{(ij)}_{\mu \nu} .
\end{eqnarray}
By substituting Eqs. (\ref{gh385}-\ref{gh390}) into Eq. (\ref{gh354}), we obtain
\begin{eqnarray}\nonumber
\mathcal{M}^\mathrm{ref}(\brr,t)
 & =  & \frac{e^{i ( k_0 z'' - \omega_0 t)}}{(2 \pi)^{3/2}} \sum_{\mu, \nu =1}^2 \ell_{\mu \nu}  r_\nu(\bk_0) \int
\left[ \mathbf{E}_{\mu \nu} +  \frac{q_i}{k_0} \mathbf{E}^{(i)}_{\mu \nu} + \frac{1}{2 k_0^2} q_i q_j \mathbf{E}^{(ij)}_{\mu \nu} \right] \\
\label{gh400}
& &  \times \exp  \left( \frac{q_i}{k_0} \upsilon_\nu^i + \frac{1}{2 k_0^2}q_i q_j  \mathcal{U}_\nu^{ij} \right)  F(D \bq) \exp[i D \bq\cdot X \bx'' - i
\delta \omega(\bk_0, \bq)t] \di^3 q .
\end{eqnarray}
%
%
Since, from Eq. (\ref{b170}) we know that
\begin{eqnarray}\label{gh410}
F(\bp) = \frac{(2 \pi)^{1/2} L}{k_0} \exp \left[
-\frac{L}{2 k_0} (p_1^2 + p_2^2) \right] \delta \left( p_3 +
\frac{p_1^2 + p_2^2 }{2 k_0} \right),
\end{eqnarray}
it is convenient to perform a change of integration variables into  Eq. (\ref{gh400}) by putting $\bq \rightarrow \bp = D \bq \Rightarrow \bq = D^T \bp$, thus obtaining
\begin{eqnarray}\nonumber
\mathcal{M}^\mathrm{ref}(\brr,t)
 & =  &  \frac{e^{i ( k_0 z'' - \omega_0 t)}}{(2 \pi)^{3/2}} \sum_{\mu, \nu =1}^2 \ell_{\mu \nu}  r_\nu(\bk_0) \int
\Bigl[ \mathbf{E}_{\mu \nu} +  \frac{p_\alpha D_{\alpha i} + p_3 D_{3 i}}{k_0} \mathbf{E}^{(i)}_{\mu \nu} \Bigr. \\ \nonumber
& & \Bigl. + \frac{1}{2 k_0^2} (p_\alpha D_{\alpha i} + p_3 D_{3 i})(p_\beta D_{\beta j} + p_3 D_{3 j}) \mathbf{E}^{(ij)}_{\mu \nu} \Bigr] \\
 \nonumber
& &  \times \exp  \left[ \frac{p_\alpha D_{\alpha i} + p_3 D_{3 i}}{k_0} \upsilon_\nu^i + \frac{1}{2 k_0^2} (p_\alpha D_{\alpha i} + p_3 D_{3 i})(p_\beta D_{\beta j} + p_3 D_{3 j})  \mathcal{U}_\nu^{ij} \right] \\ \label{gh405}
 & & \times F(\bp) \exp[i (p_\alpha X_{\alpha j} x_j'' +  p_3 X_{3 j} x_j'') - i
\delta \omega(\bk_0, D^T\bp)t] \di^3 p,
\end{eqnarray}
where we have displayed the variable $p_3$ in order to integrate with respect to it by exploiting the Dirac delta present in the expression of $F(\bp)$. Such integration yields the substitution
\begin{eqnarray}\label{gh420}
p_3 \rightarrow -\frac{1}{2 k_0} p_\alpha p_\beta \delta_{\alpha \beta},
\end{eqnarray}
everywhere in Eq. (\ref{gh405}).
In addition, from Eq. (\ref{gh420}) and Eq. (\ref{b380}), it immediately follows:
\begin{eqnarray}\label{gh430}
\omega(\bk_0, D^T \bp) \cong \frac{\omega_0}{8} \left( \frac{p_1^2 +
p_2^2}{k_0^2} \right)^2 \simeq 0,
\end{eqnarray}
where the last approximate equality follows from our choice to keep
expansion terms up to the second order in the phase terms. This result is important because it tells us that  up to the second order terms, the beam is monochromatic. Therefore, we are enabled  to use, for the  calculation of the Poynting vector  $\mathbf{S}(\brr)$, of the cycle-average theorem to obtain $\mathbf{S}(\brr) \propto \mathrm{Re} \left[ \mathbf{E}(\brr,t) \times  \mathbf{B}^*(\brr,t) \right]$; we shall use soon this result.  Within the same level of approximation, we can thus rewrite Eq. (\ref{gh405}) after integration with respect to $p_3$, as:
\begin{eqnarray}\nonumber
\mathcal{M}^\mathrm{ref}(\brr,t)
 & \simeq  & \frac{e^{i ( k_0 z'' - \omega_0 t)}}{(2 \pi)^{3/2}} \sum_{\mu, \nu =1}^2 \ell_{\mu \nu}  r_\nu(\bk_0) \int
\Bigl[ \mathbf{E}_{\mu \nu} +  \frac{p_\alpha }{k_0}D_{\alpha i} \mathbf{E}^{(i)}_{\mu \nu} \Bigr. \\ \nonumber
& & \Bigl. - \frac{1}{2 } \frac{p_\alpha}{k_0} \frac{p_\beta}{k_0} \left(\delta_{\alpha \beta} D_{3i} \mathbf{E}^{(i)}_{\mu \nu} -  D_{\alpha i}D_{\beta j}  \mathbf{E}^{(ij)}_{\mu \nu} \right)\Bigr] \frac{(2 \pi)^{1/2} L}{k_0} \\ \nonumber
& &  \times \exp  \left\{ \frac{p_\alpha}{k_0} \left( D_{\alpha i} \upsilon_\nu^i + i k_0 X_{\alpha i} {x''}^i \right) \right. \\ \label{gh440}
& & \left.  + \frac{1}{2 k_0^2} p_\alpha p_\beta \left[ D_{\alpha i}D_{\beta j} \mathcal{U}_\nu^{ij} - \delta_{\alpha \beta} \left( D_{3 i} \upsilon_\nu^i + i k_0 z'' + k_0 L\right)   \right] \right\} \di^2 p,
\end{eqnarray}
where we have used Eq. (\ref{b490}) to write $X_{3 i} {x''}^i = \delta_{3 i}(-1)^{\delta_{1 i}}{x''}^i = {x''}^3 \equiv z''$.
If we introduce the dimensionless variables $\{P_\alpha \equiv p_\alpha / k_0 \}_{\alpha =1}^2$, then we can rewrite Eq. (\ref{gh440}) as
\begin{eqnarray}\nonumber
\mathcal{M}^\mathrm{ref}(\brr,t)
 & \simeq  & \frac{k_0 L }{2 \pi} e^{i ( k_0 z'' - \omega_0 t)} \sum_{\mu, \nu =1}^2 \ell_{\mu \nu}  r_\nu(\bk_0) \int
\Bigl[ \mathbf{E}_{\mu \nu} +  P_\alpha D_{\alpha i} \mathbf{E}^{(i)}_{\mu \nu} \Bigr. \\ \nonumber
& & \Bigl. - \frac{1}{2 } P_\alpha P_\beta \left(\delta_{\alpha \beta} D_{3i} \mathbf{E}^{(i)}_{\mu \nu} -  D_{\alpha i}D_{\beta j}  \mathbf{E}^{(ij)}_{\mu \nu} \right)\Bigr] \\ \label{gh450}
& &  \times \exp  \left[ - \frac{1}{2} P_\alpha  (B^{-1}_\nu)_{\alpha \beta} P_\beta +  P_\alpha b_{\nu \alpha} \right]\di^2 P,
\end{eqnarray}
where we have defined the two two-dimensional vectors $\mathbf{b}_\nu $ $(\nu = 1,2)$, and the two \emph{symmetric} $2 \times 2$ matrices $B_\nu$, whose elements are
\begin{eqnarray}\nonumber
b_{\nu \alpha} \equiv (\mathbf{b}_\nu)_\alpha & =  &  i k_0 \left[ D (\bvarphi_\nu - i \brho_\nu)  + X {\bx''} \right], \\ \label{gh460}
(B^{-1}_\nu)_{\alpha \beta} &\equiv &  \delta_{\alpha \beta} \left( b_{\nu 3} + k_0 L\right)-ik_0^2 [D (\mathcal{F}_\nu - i \mathcal{R}_\nu   )D^T]_{\alpha \beta}    ,
\end{eqnarray}
and $\alpha, \beta \in \{1,2\}$. Note that in the case of geometric reflection, the expressions above reduces to $\mathbf{b}_\nu =  i k_0 X \bx''$, and $(B^{-1}_\nu)_{\alpha \beta} =   i k_0 \delta_{\alpha \beta} \left(z'' - i L \right)$, respectively. Moreover, it is important to note that the off-diagonal terms of the matrix $B$, which change the profile of the beam, arise because of the second order terms $\sim \mathcal{F}_\nu - i \mathcal{R}_\nu $; while fist  order terms $\sim \bvarphi_\nu - i \brho_\nu$ simply produce a shift.
Equation (\ref{gh450}) can be analytically integrated by using the following three basic formulas for Gaussian integrals \cite{SchulmanBook}:
\begin{eqnarray}\nonumber
G_\nu &  = &  \int  \exp \left[ -\frac{1}{2} \sum_{ \alpha, \beta =1 }^2
P_\alpha (B^{-1}_\nu)_{\alpha \beta} P_\beta + \sum_{ \alpha =1 }^2
P_\alpha b_{\nu \alpha} \right]  \di^2 P\\ \label{gh462}
&  = &  2 \pi ( \det B_\nu)^{1/2} \exp \left[ \frac{1}{2} \sum_{ \alpha,
\beta =1 }^2 b_{\nu \alpha} (B_\nu)_{\alpha \beta} b_{\nu \beta }\right] ,
\end{eqnarray}
\begin{eqnarray}\nonumber
G_{\nu,\sigma} &  = &  \int P_\sigma \exp \left[ -\frac{1}{2} \sum_{ \alpha, \beta =1 }^2
P_\alpha (B^{-1}_\nu)_{\alpha \beta} P_\beta + \sum_{ \alpha =1 }^2
P_\alpha b_{\nu \alpha} \right]  \di^2 P\\ \label{gh463}
&  = &  \frac{\partial G_\nu}{\partial b_{\nu \sigma}} = G_\nu \left(b_{\nu \alpha} (B_\nu)_{\alpha \sigma} \right) = G_\nu \left(B_\nu \bb_\nu \right)_\sigma ,
\end{eqnarray}
\begin{eqnarray}\nonumber
G_{\nu,\sigma \tau} &  = &  \int P_\sigma P_\tau \exp \left[ -\frac{1}{2} \sum_{ \alpha, \beta =1 }^2
P_\alpha (B^{-1}_\nu)_{\alpha \beta} P_\beta + \sum_{ \alpha =1 }^2
P_\alpha b_{\nu \alpha} \right] \di^2 P\\ \nonumber
&  = &  \frac{\partial^2 G_\nu}{\partial b_\sigma  \partial b_\tau}  = G_\nu \left[(B_\nu)_{\sigma \tau} + (b_{\nu \alpha} (B_\nu)_{\alpha \sigma})( b_{\nu \beta} (B_\nu)_{\beta \tau}) \right]
\\ \label{gh464}
&  = & G_\nu \left[B_\nu + B_\nu (\bb_\nu \bb_\nu) B_\nu \right]_{\sigma \tau}
.
\end{eqnarray}
By using the equations above into Eq. (\ref{gh450}), we easily obtain our final result:
\begin{eqnarray}\nonumber
\mathcal{M}^\mathrm{ref}(\brr,t)
 & \simeq  & \frac{k_0 L }{2 \pi} e^{i ( k_0 z'' - \omega_0 t)} \sum_{\mu, \nu =1}^2 \ell_{\mu \nu}  r_\nu(\bk_0) \Bigl\{
 G_\nu \mathbf{E}_{\mu \nu} +  G_{\nu,\alpha}  D_{\alpha i} \mathbf{E}^{(i)}_{\mu \nu} \Bigr. \\ \nonumber
& & \Bigl. - \frac{1}{2 } G_{\nu, \alpha  \beta} \left( \delta_{\alpha \beta} D_{3i}  \mathbf{E}^{(i)}_{\mu \nu} -  D_{\alpha i}D_{\beta j}  \mathbf{E}^{(ij)}_{\mu \nu} \right) \Bigr\} \\ \nonumber
\\ \nonumber
%
%
 %
 & =  & \frac{k_0 L }{2 \pi} e^{i ( k_0 z'' - \omega_0 t)} \sum_{\mu, \nu =1}^2 \ell_{\mu \nu}  r_\nu(\bk_0)  G_\nu  \Bigl\{
\mathbf{E}_{\mu \nu} +  (B_\nu \bb_\nu)_{\alpha}  D_{\alpha i} \mathbf{E}^{(i)}_{\mu \nu} \Bigr. \\ \label{gh470}
& & \Bigl. - \frac{1}{2 } \left[B_\nu + B_\nu (\bb_\nu \bb_\nu) B_\nu \right]_{\alpha \beta} \left( \delta_{\alpha \beta} D_{3i}  \mathbf{E}^{(i)}_{\mu \nu} -  D_{\alpha i}D_{\beta j}  \mathbf{E}^{(ij)}_{\mu \nu} \right) \Bigr\},
%
%
 %
%
%
%
%
\end{eqnarray}
with $i,j \in \{1,2,3\}$.
From this equation we can calculate analytically both the electric and magnetic fields and, consequently, the Poynting vector. From the latter we have automatically the energy flux which is the quantity experimentally accessible.
\subsection{First order Taylor expansion}
\begin{align}\nonumber
\psi(k_{0x} + q_x,k_{0y}+q_y,k_{0z}+q_z ) & \cong \psi(k_{0x},k_{0y},k_{0z} ) +
 \left. \left( q_x \frac{\partial \psi}{\partial k_x} +  q_y  \frac{\partial \psi}{\partial k_y} +  q_z  \frac{\partial \psi}{\partial k_z}\right)   \right|_{\bk = \bk_0} + \ldots \\ \label{g30}
 & = \psi(k_{0x},k_{0y},k_{0z} ) +
 \left. \bq \cdot \nabla_\bk \psi   \right|_{\bk = \bk_0} + \ldots
\end{align}
where
\begin{align}\nonumber
\nabla_\bk & = \ve_1 \frac{\partial }{\partial k_x} +  \ve_2  \frac{\partial }{\partial k_y} +  \ve_3  \frac{\partial }{\partial k_z} \\ \nonumber
 & = \ve_\theta \frac{1}{k} \frac{\partial }{\partial \theta} +  \ve_\phi   \frac{1}{k \sin \theta} \frac{\partial }{\partial \phi} +  \ve_k  \frac{\partial }{\partial k}
\\
 & = \ve_1(\bk) \frac{1}{k} \frac{\partial }{\partial \theta} +  \ve_2(\bk)   \frac{1}{k \sin \theta} \frac{\partial }{\partial \phi} +  \ve_3 (\bk)  \frac{\partial }{\partial k}.
\end{align}
If $\psi(k_x,k_y,k_z) = \psi(\theta)$ then
\begin{align}\nonumber
\psi(\bk_0 + \bq ) & \cong \psi(\bk_0) +
 \left. \frac{1}{k_0} \frac{ \partial \psi }{\partial \theta} \right|_{\theta = \theta_0} \ve_1(\bk_0) \cdot \bq + \ldots \\ \label{g30}
 & = \psi(\bk_0) +
 \left. \frac{1}{k_0} \left( q_x \cos \theta_0 - q_z \sin \theta_0 \right) \frac{ \partial \psi }{\partial \theta} \right|_{\theta = \theta_0} + \ldots
\end{align}
\begin{align}\nonumber
\mathbf{E}^\mathrm{ref}(\bx'',t)
 \cong   & \frac{k_0 L }{2 \pi} e^{i ( k_0 z'' - \omega_0 t)}  r_1(\bk_0) \Biggl\{ \Bigr. \\ \nonumber
 &
 \quad \; \be''_1 \left[ f_1 G_1 + y'' \cot \theta_0 \frac{f_2}{z'' - i L} \left( G_1 + \rho G_2\right)\right]
 \\ \nonumber
 &  + \be''_2 \left[ f_2 \rho G_2 - y'' \cot \theta_0 \frac{f_1}{z'' - i L} \left( G_1 + \rho G_2\right)\right]   \\ \nonumber
 &  - \be''_3 \left[\frac{ f_1 G_1 (x'' - v_1^0) -y'' f_2 \rho G_2}{z'' - i L}   \right] \Biggl. \Biggr\}
\end{align}
\begin{align}\nonumber
\mathbf{B}^\mathrm{ref}(\bx'',t)
 \cong  & - \frac{k_0 L }{2 \pi c} e^{i ( k_0 z'' - \omega_0 t)}  r_1(\bk_0)  \Biggl\{ \Bigr. \\ \nonumber
 &
 \quad \; \be''_1 \left[ f_2 \rho G_2 - y'' \cot \theta_0 \frac{f_1}{z'' - i L} \left( G_1 + \rho G_2\right)\right]
 \\ \nonumber
 &  - \be''_2 \left[ f_1 G_1 + y'' \cot \theta_0 \frac{f_2}{z'' - i L} \left( G_1 + \rho G_2\right)\right]  \\ \nonumber
 &  - \be''_3 \left[\frac{ f_2 \rho G_2 (x'' - v_2^0) -y'' f_1  G_1}{z'' - i L}   \right] \Biggl. \Biggr\}
\end{align}
%
\appendix
\section{Complete bases in $\mathbb{R}^3$}
Let $K \equiv (Oxyz)$ be a  Cartesian reference frame, and
let $\bk$ denotes a  vector pointing along the direction
${\bm k}: \, \bk = k {\bm k}$, where $k \equiv
|\bk|$ and $|{\bm k}|=1$. Moreover, let $\{ \be_x, \be_y,
\be_z\}$ be three mutually orthogonal unit vectors pointing along
the axes $x,y$ and $z$, respectively.
Given $\bk$ and $\be_z$, we can built a
complete basis $\{ \be_{1}(\bk) , \be_{2}(\bk) ,
\be_{3}(\bk) \}$ in $\mathbb{R}^3$ by using the following recipe:
\begin{align}\label{a71}
 \be_{1}(\bk)  &=  \frac{(\be_z \times \bk)\times \bk}
 {|(\be_z \times \bk)\times \bk|},
 \\ \label{a81}
 \be_{2}(\bk)  &=
 \frac{\be_z \times \bk}{|\be_z \times
 \bk|},
  \\ \label{a91}
 \be_{3}(\bk)  &=
 \frac{\bk}{|
 \bk|},
\end{align}
%
where ``$\, \times \,$'' denotes the ordinary vector product in
$\mathbb{R}^3$:
\begin{eqnarray}\label{a25}
(\ba \times \bb)_i = \epsilon_{ijk} a_j b_k, \qquad (i,j,k \in
\{1,2,3 \}),
\end{eqnarray}
and $ \epsilon_{ijk}$ is the completely antisymmetric
Levi-Civita tensor, such that $ \epsilon_{ijk} = + 1$ or $-1$
according to whether the indices $i,j,k$ are an even or odd
permutation of the ordered set $\{ 1,2,3\}$, and $ \epsilon_{ijk}
= 0$ when at least two indices are equal.
By definition, the three real-valued unit vectors $\{\be_{i} (\bk)
\}_{i=1}^3$ form an orthogonal
%
\begin{align}\label{a100}
 \be_{i}(\bk) \cdot  \be_{j}(\bk) &= \delta_{ij},
  \\ \label{a110}
   \be_{i}(\bk) \times  \be_{j}(\bk) &= \sum_{k=1}^3
 \epsilon_{ijk}\be_{k}(\bk),
 \end{align}
%
$(i,j \in \{1,2,3 \})$ and complete basis in $\mathbb{R}^3$:
\begin{eqnarray}\label{a80}
\sum_{i=1}^3  \be_{i}(\bk) \be_{i}(\bk) =  I ,
\end{eqnarray}
where $ I $ denotes the $3 \times 3$ identity matrix, and we have used the symbol $\ba \bb$
to denote the \emph{diadic} product of the two vectors $\ba$ and $\bb$, respectively, that generates a matrix whose elements are:
$(\ba \bb)_{ij} = a_i b_j$, $(i,j=1,2,3)$.
 Since $\be_{3}(\bk) \be_{3}(\bk) = \bk
 \bk/k^2$, from Eq. (\ref{a80}) it readily follows that
\begin{align}\label{a90}
\sum_{\lambda=1}^2  \be_{\lambda}(\bk) \be_{\lambda}(\bk) & =  I  -
\frac{\bk \bk }{k^2} \\
& \equiv  I  - \mathcal{L}(\bk) \\
& \equiv \mathcal{T}(\bk),
\end{align}
where with $\mathcal{L}(\bk)$ and $\mathcal{T}(\bk)$ we denoted the longitudinal and the transverse projectors, respectively.
The main advantage of using the definitions (\ref{a71})-(\ref{a91}) is that
if we write $\bk$ in
spherical polar coordinates $(k,\theta, \phi)$ as
\begin{eqnarray}\label{a95}
\bk = k(\be_1 \sin \theta \cos \phi + \be_2 \sin
\theta  \sin \phi  + \be_3 \cos \theta ),
\end{eqnarray}
then it is easy to see  that
%
\begin{align}\label{a51}
 \be_{1}(\bk)  &=  \be_1 \cos \theta  \cos \phi  + \be_2 \cos \theta
\sin \phi  - \be_3 \sin \theta,
 \\ \label{a61}  \be_{2}(\bk)
&=  -\be_1 \sin \phi  + \be_2 \cos  \phi ,
 \\ \label{a62}  \be_{3}(\bk)  &= \be_1 \sin \theta  \cos \phi  + \be_2 \sin \theta
\sin \phi  + \be_3 \cos \theta.
\end{align}
%
Now, from Eqs. (\ref{a51})-(\ref{a62}) it immediately follows that the basis $\{ \be_{1}(\bk) , \be_{2}(\bk) , \be_{3}(\bk) \}$ coincides with the  spherical basis $\{ \be_\theta , \be_\phi , \be_{k} \}$, namely
\begin{align}\label{a200}
 \be_{1}(\bk) &= \; \frac{1}{k } \frac{\partial
\bk  }{\partial \theta }   \equiv  \be_{\theta},
 \\ \label{a210}  \be_{2}(\bk) &= \;\frac{1}{k  \sin
\theta } \frac{\partial \bk   }{\partial \phi }
\equiv \be_{\phi} ,
 \\ \label{a220}  \be_{3}(\bk)  &=  \; \frac{\partial  \bk
}{\partial k }  \equiv \be_{k}.
\end{align}
Finally, we stress that the choice of the recipe (\ref{a71}-\ref{a91}) is determined by the necessity of having two unit vectors parallel and orthogonal to the plane of incidence common to $\bk$ and $\be_z$.
%
%
%
\section{Fourier transform and bases change}
In this appendix we shortly review a few elementary facts about
Fourier transform and basis changes. To begin with, let us study
how the Fourier transform $F(p_1,p_2,p_3)$  of a given smooth
function $f(x_1,x_2,x_3)$ changes, when we pass from the initial
set of coordinates  $\{ x_1,x_2,x_3 \}$ to a new one $\{
u_1,u_2,u_3 \}$, via a \emph{linear} transformation.
By definition of Fourier transform, the two functions
$f(x_1,x_2,x_3)$ and $F(p_1,p_2,p_3)$ are related by the formulas
%
\begin{eqnarray}\label{fo10}
 f(x_1,x_2,x_3)  &= &\frac{1}{(2 \pi)^{3/2}} \int
F(p_1,p_2,p_3) e^{i(x_1 p_1 + x_2 p_2 + x_3 p_3)} \mathrm{d} p_1
\, \mathrm{d} p_2 \, \mathrm{d} p_3 ,\\\label{fo15}
F(p_1,p_2,p_3) & =&  \frac{1}{(2 \pi)^{3/2}} \int f(x_1,x_2,x_3)
e^{-i(x_1 p_1 + x_2 p_2 + x_3 p_3)} \mathrm{d} x_1 \, \mathrm{d}
x_2 \, \mathrm{d} x_3,
\end{eqnarray}
%
with $x_i \in \mathbb{R}$, $p_i \in \mathbb{R}$, $(i=1,2,3)$, and
we use the convention that all integrals are calculated in the
interval $(- \infty, \infty)$, unless otherwise stated. The
validity of the equations above can be easily checked by
substituting Eq. (\ref{fo15}) into Eq. (\ref{fo10}), interchanging
the order of integration,  and then using the following integral
representation for the Dirac delta function:
\begin{eqnarray}\label{fo30}
 \delta^{(3)}(\bx - \bx')  = \frac{1}{(2 \pi)^{3}} \int e^{i \bp \cdot (\bx - \bx')} \di^3 x,
\end{eqnarray}
where $\bx \equiv (x_1,x_2,x_3)$, $\bp \equiv (p_1, p_2, p_3)$,
$\di^3 x \equiv \di x_1 \di x_2 \di x_3$, and the dot ``$\; \cdot
\;$'' denotes the ordinary scalar product in $\mathbb{R}^3$, e.g.,
$\bx \cdot \bp = x_i p_i$ and, from now on, we sum over repeated
indices unless otherwise stated.
Now, let $\{u_1, u_2, u_3 \}$ be a new set of coordinates related
to the old ones  $\{x_1, x_2, x_3 \}$ via a non-homogeneous
\emph{orthogonal} transformation:
\begin{eqnarray}\label{fo40}
x_i = D_{ij} u_j + a_i \quad \Leftrightarrow \quad \bx = D \bu +
\ba,
\end{eqnarray}
where, by hypothesis, $D D^T = I$. Under this transformation, the
infinitesimal volume $\di^3 \, x$ becomes
\begin{eqnarray}\label{fo50}
\di x_1 \, \di x_2 \, \di x_3 \,=  \left( \det J \right)  \di u_1
\, \di u_2 \, \di u_3,
\end{eqnarray}
where $J: \, J_{ij} = \partial x_i / \partial u_j = D_{ij}$, $(i,j
=1,2,3)$, denotes the Jacobian of the transformation. If we use
Eq. (\ref{fo40}) in both sides of Eq. (\ref{fo10}), we easily obtain
\begin{eqnarray}\nonumber
 f(D
\bu + \ba)  &= &\frac{1}{(2 \pi)^{3/2}} \int F(\bp) e^{i(D \bu +
\ba)\cdot \bp} \mathrm{d}^3  p, \\ \nonumber
 &= &\frac{1}{(2
\pi)^{3/2}} \int  F(\bp) e^{i(\bu \cdot D^T
 \bp +     \ba \cdot \bp )} \mathrm{d}^3  p, \\
\nonumber
 &= &\frac{1}{(2
\pi)^{3/2}} \int F (D\bk)  e^{i(\bu \cdot
 \bk +     \ba \cdot  D \bk )} \mathrm{d}^3  k, \\
\label{fo60}
 &= &\frac{1}{(2
\pi)^{3/2}} \int \left[ F(D\bk ) e^{i \ba \cdot  D \bk } \right]
e^{i\bu \cdot
 \bk } \mathrm{d}^3  k,
\end{eqnarray}
where  $|\det J| = |\det D| =1$, and
\begin{eqnarray}\nonumber
D\bu \cdot \bp  &= &(D \bu)_i p_i = D_{ij} u_j p_i = u_j D^T_{ji}
p_i = \bu \cdot D^T \bp.
\end{eqnarray}
Moreover, in the third line of Eq. (\ref{fo60}) we have changed
integration variables passing from $\bp$ to  $\bk: \bk \equiv D^T
\bp$ with, once again, $|\det J| = |\det D | =1$. On the other
hand, if we define
\begin{eqnarray}\label{fo70}
 f(D
\bu + \ba)  \equiv g(\bu),
\end{eqnarray}
then, by definition of Fourier transform as given in Eq.
(\ref{fo10}), we can write
\begin{eqnarray}\label{fo80}
g(\bu) =\frac{1}{(2 \pi)^{3/2}} \int G(\bk) e^{i \bu \cdot \bk}
\mathrm{d}^3  k.
\end{eqnarray}
At this point, it is enough to equate the right sides of Eqs.
(\ref{fo60}-\ref{fo80}) to find the sought relation:
\begin{equation}\label{fo90}
 G(\bk) =  F(D\bk ) e^{i \ba \cdot  D \bk } .
\end{equation}
%
%
\section{Gaussian beams and bases change}
Let $K \equiv (Oxyz)$ be a given Cartesian reference frame, and
let $\brr_0$ denotes a given vector pointing along the direction
${\bm r}_0: \, \brr_0 = r_0 {\bm r}_0$, where $r_0 \equiv
|\brr_0|$ and $|{\bm r}_0|=1$. Moreover, let $\{ \be_1, \be_2,
\be_3\} = \{ \be_x, \be_y,
\be_z\}$ be three mutually orthogonal unit vectors pointing along
the axes $x,y$ and $z$, respectively. The two vectors ${\bm r}_0$
and $\be_3$ determines a plane, say, \emph{the plane of
incidence}. If we think of $\vkappa_0 \equiv - {\bm r}_0$ as a unit vector pointing
along the axis $z'$ of a new frame $K' \equiv (O'x'y'z')$ centered
in $\brr_0$, as shown in Fig. C1, we are free to choose the directions of the axes $x'$
and $y'$ around the axis $z'$.
 By convention, we choose $x'$ on
the plane of incidence and $y'$ orthogonal to it. Then, if we
denote with $\{ \be_1', \be_2', \be_3'\}$  three unit vectors
pointing along the axes $x',y'$ and $z'$, respectively, the
convention above is automatically satisfied by choosing, according to the
results of Appendix A, $\{ \be_1', \be_2', \be_3'\} =
 \{ \be_1(\vkappa_0), \be_2(\vkappa_0), \be_3(\vkappa_0)\}$.

Now, let $P$ an arbitrary point of coordinates
$\overrightarrow{OP} \equiv \brr \doteq (x_1,x_2,x_3) \equiv \brr$
in $K$, and coordinates $\overrightarrow{O'P} \equiv \brr' \doteq
(x'_1,x'_2,x'_3) \equiv \bx'$ in $K'$, where the symbol
``$\doteq$'' stands for: ``is represented by''. Moreover, the
coordinates of the origin $O'$ of $K'$, with respect to  $K$, are
equal to $\overrightarrow{OO'} \equiv \brr_0 \doteq
(x_{01},x_{02},x_{03}) \equiv \brr_0$, by definition. Note that,
in order to avoid confusion, we use the  symbols $\brr, \brr_0$ to
denote \emph{both} the vectors and their representation in the
frame $K$, while we use the symbol $\bx'$ for the representation
of the vector $\brr'$ in the frame $K'$.
\begin{flushright}
\begin{figure}[!hr]
\includegraphics[angle=0,width=9truecm]{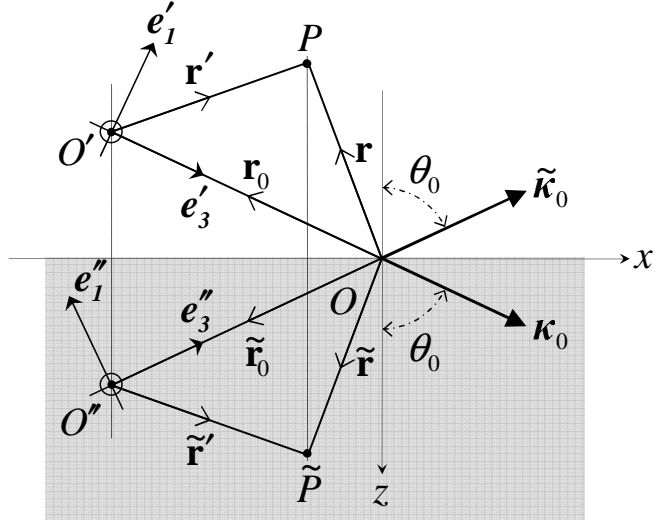}
\caption{\label{fig:C1}
Coordinate system definition. The two vectors $\be_2'$ and $\be_2''$ are directed towards the reader, parallel to  $\be_y$.}
\end{figure}
\end{flushright}
Since,  trivially,
\begin{eqnarray}\label{b100}
\overrightarrow{OO'} + \overrightarrow{O'P} = \overrightarrow{OP}
\quad \Leftrightarrow \quad \brr_0 + \brr' = \brr,
\end{eqnarray}
then, we can  write
%
\begin{eqnarray}\nonumber
x_i' & \equiv & \be_i' \cdot \brr'
 \\  \label{b105}
& = & \be_i' \cdot (\brr - \brr_0)
 \\ \nonumber
& = & \sum_{j=1}^3 \be_i' \cdot \be_j (x_j - x_{0j})
 \\ \label{b110}
& \equiv & \sum_{j=1}^3 D_{ij} (x_j - x_{0j}),
\end{eqnarray}
%
where in the last line we have defined $D_{ij} \equiv \be_i' \cdot
\be_j$. If we define
\begin{eqnarray}\label{b120}
a_i \equiv - D_{ij}  x_{0j} = - \be_i' \cdot \brr_0 = r_0 \be_i'
\cdot \vkappa_0 = r_0 \delta_{i3},
\end{eqnarray}
then we can rewrite in compact form
\begin{eqnarray}\label{b122}
x_i' = D_{ij} x_j + a_i = \be_i' \cdot \brr + r_0 \delta_{i3},
\end{eqnarray}
or, equivalently,
\begin{eqnarray}\label{b124}
\bx' = D (\brr - \brr_0) \equiv D \brr + \ba,
\end{eqnarray}
where, in analogy with Eq. (\ref{fo40}), we have defined $\ba
\equiv - D \brr_0$.
It is easy to prove that $D$ is orthogonal, that is $D D^T = I$:
\begin{eqnarray}\nonumber
(D D^T)_{ij} = D_{ik}D_{jk} &= &(\be_i' \cdot \be_k)( \be_j' \cdot
\be_k)\\ \nonumber
& = & (\be_i')_k ( \be_j' )_k \\ \nonumber
& = &\be_i' \cdot \be_j' \\ \label{b130}
&= &\delta_{ij} = ( I )_{ij},
\end{eqnarray}
where in the second line we have used the definition of
coordinates of the vectors $\be_i'$ with respect to the basis
$\be_k$: $(\be_i')_k \equiv \be_k \cdot \be_i' = \be_i' \cdot
\be_k$.

Now, let us consider a Gaussian beam propagating along the
positive direction of the axis $z'$, at wavelength $\lambda_0$
with a central waist $w_0 = w(z'=0)$ located at $\brr_0$. Its
functional shape is given by
\begin{eqnarray}\label{b140}
\psi(x',y',z',t) = \exp [i k_0 (z' - ct)] f(x',y',z'),
\end{eqnarray}
where $k_0 \equiv 2 \pi  / \lambda_0$ and
\begin{eqnarray}\label{b150}
 f(x',y',z')= \frac{- i
L}{z' - i L} \exp \left[ i \frac{k_0}{2}\left( \frac{x'^{\, 2} +
y'^{\, 2}}{z'-i L} \right)\right],
\end{eqnarray}
is a solution of the paraxial wave equation
\begin{eqnarray}\label{b160}
\left( \frac{\partial^2}{\partial x^2} +
\frac{\partial^2}{\partial y^2} + 2 i k_0 \frac{\partial}{\partial
z} \right)f(x,y,z) = 0,
\end{eqnarray}
with $L = k_0 w_0^2/2$ (the Raleigh range of the beam). Let us
calculate the Fourier transform of $f(x',y',z')$ as
\begin{eqnarray}\nonumber
F(p_1,p_2,p_3) & = & \frac{1}{(2 \pi)^{3/2}} \int
f(x',y',z')e^{-i(x' p_1 + y' p_2 + z' p_3)}\mathrm{d} x'
\mathrm{d} y' \mathrm{d} z'
\\ \label{b170}
 & = &\frac{(2 \pi)^{1/2} L}{  k_0}\exp\left[ -\frac{L}{2
k_0}\left( p_1^2 + p_2^2 \right) \right]\delta \left( p_3 +
\frac{p_1^2 + p_2^2}{2 k_0} \right),
\end{eqnarray}
where we have used the elementary Gaussian formula
\begin{align}\nonumber
\int_{-\infty}^\infty \exp \left[ \left(a_1 + i b_1  \right) x  \right.  &+
 \left. \left(a_2 + i b_2  \right) x^2
 \right] \mathrm{d} x
 \\ \label{b180}
&  = \left( \frac{\pi}{-a_2 - i b_2} \right)^{1/2} \exp \left[
-\frac{(a_1 + i b_1)^2}{4(a_2 + i b_2)} \right], \qquad (a_2 < 0),
\end{align}
with $\{a_i,b_i\}_{i=1}^2 \in \mathbb{R}$, to calculate
\begin{align}\nonumber
I(u,z') & \equiv  \int_{-\infty}^\infty \exp \left( i u x + i
\frac{k_0}{2} \frac{x^2}{z' - i L} \right) \mathrm{d} x \\
\label{b190}
& =  \left[ \frac{2 \pi i}{k_0} (z' - i L)\right]^{1/2} \exp \left[
-\frac{i u^2}{2 k_0} ( z' - i L ) \right].
\end{align}
Next, what we want to do is to express $f(x',y',z')=f(\bx')$ in
the frame $K$, that is we want to find the function $g(\brr)$ such
that $g(\brr) = f\left(\bx' = D \brr - D \brr_0 \right)$. From Eq.
(\ref{b105}) it immediately follows that
\begin{eqnarray} \nonumber
g(\brr )&= &\frac{- i L}{\be_3' \cdot (\brr - \brr_0) - i L} \\
\nonumber
& & \times \exp \left\{ i \frac{k_0}{2} \frac{[\be_1' \cdot (\brr
- \brr_0)]^{\, 2} + [\be_2' \cdot (\brr - \brr_0)]^{\,
2}}{\be_3' \cdot (\brr - \brr_0)-i L} \right\} \\
\nonumber
&= &\frac{- i L}{\be_3' \cdot \brr +r_0 -i L} \\
\label{b200}
 & & \times \exp \left[ i \frac{k_0}{2} \frac{(\be_1'
\cdot \brr )^{\, 2} + (\be_2' \cdot \brr )^{\, 2}}{\be_3' \cdot
\brr +r_0 -i L} \right],
\end{eqnarray}
and that
\begin{eqnarray}\nonumber
 \exp [i k_0 (z' - ct)] & = & \exp [i k_0 \be_3' \cdot (\brr - \brr_0) - i k_0c
 t]\\ \label{b210}
 & = & \exp [i \bk_0 \cdot (\brr - \brr_0) - i \omega_0
 t],
\end{eqnarray}
where, by definition, $\bk_0 = k_0 \be_3'$, and $\omega_0 \equiv
k_0 c$. Thus, we can rewrite our initial Gaussian beam
$\psi(\bx',t)$ in the frame $K$ as $\psi(\bx',t) \equiv \Psi(\brr,
t)$, where
\begin{eqnarray} \label{b220}
 \Psi(\brr, t)= g(\brr) \exp[i \bk_0 \cdot (\brr - \brr_0) - \omega_0
 t].
\end{eqnarray}
It is well know that this field is \emph{not} a solution of the
Maxwell wave equation, that is
\begin{eqnarray}\label{b230}
 \left( \frac{\partial^2}{\partial
x^2} + \frac{\partial^2}{\partial y^2} +
\frac{\partial^2}{\partial z^2}- \frac{1}{c^2}
\frac{\partial^2}{\partial t^2} \right) \Psi(\brr, t)  = \Box
 \Psi(\brr, t) \neq 0.
\end{eqnarray}
However, we can seek a field $\Psi_0(\brr, t)$ such that $\Box
 \Psi_0(\brr, t) = 0, \, \forall t>0$, and
\begin{eqnarray}\label{b235}
   \Psi_0(\brr, t=0) = \Psi(\brr,
 t=0).
 \end{eqnarray}
In order to determine $\Psi_0(\brr, t)$, we begin by writing it as
 a plane wave expansion of the form
\begin{eqnarray}\label{b240}
\Psi_0(\brr, t) = \frac{1}{(2 \pi)^{3/2}} \int a(\bk) \exp[i(\bk
\cdot \brr - \omega t)] \di^3 k
\end{eqnarray}
where $\omega \equiv |\bk| c$, and the amplitudes $a(\bk)$ have to
be found.  Then, we perform the following change of variables:
$\bk \rightarrow \bq$, where
\begin{eqnarray}\label{b250}
\bk = \bk_0 + (\bk - \bk_0) \equiv \bk_0 + \bq, \qquad \di^3 k =
\di^3 q.
\end{eqnarray}
Thus, Eq. (\ref{b240}) becomes
\begin{eqnarray}\nonumber
\Psi_0(\brr, t) & = & \frac{1}{(2 \pi)^{3/2}}  \int  a(\bk_0 +
\bq)
\exp[i (\bk_0 + \bq) \cdot \brr - i\omega( \bk_0 + \bq) t] \di^3 q \\
\label{b260}
& = & \frac{e^{i(\bk_0 \cdot \brr - \omega_0 t) }}{(2 \pi)^{3/2}}
\int a(\bk_0 + \bq) \exp[ i\bq \cdot \brr - i \delta
\omega(\bk_0,\bq) t] \di^3 q,
\end{eqnarray}
where we have defined
\begin{eqnarray} \nonumber
\delta \omega(\bk_0,\bq)   & \equiv &   \omega(\bk_0 + \bq) -
\omega_0
\\ \nonumber
&  = &  c k_0 \left( 1 + 2 \vkappa_0 \cdot \frac{\bq}{k_0} +
 \frac{q^2}{k_0^2}\right)^{1/2} - \omega_0
\\
\label{b270}
 & \simeq &
\omega_0 \left\{ \vkappa_0 \cdot \frac{\bq}{k_0} + \frac{1}{2}
\left[ \frac{q^2}{k_0^2} - \left( \vkappa_0 \cdot \frac{\bq}{k_0}
\right)^{2}\right] \left( 1 - \vkappa_0 \cdot \frac{\bq}{k_0}
\right) + \ldots \right\}.
\end{eqnarray}
with  $\vkappa_0 = \bk_0/k_0 = \be'_3$, and $q \equiv |\bq|$.  The
approximate equality in the last line of Eq. (\ref{b270}) holds in
the limit $|\bq|/{k_0} \ll 1$. At $t=0$ Eq. (\ref{b260}) can be
written as
\begin{eqnarray}\label{b280}
\Psi_0(\brr, t=0) = \frac{e^{i \bk_0 \cdot \brr   }}{(2
\pi)^{3/2}} \int a(\bk_0 + \bq) \exp( i\bq \cdot \brr ) \di^3 q.
\end{eqnarray}
By remembering Eqs. (\ref{b235}) and (\ref{b220}), and simplifying
the common term $ \exp(i \bk_0 \cdot \brr )$ on both sides, we can
rewrite Eq. (\ref{b280}) as
\begin{eqnarray}\label{b290}
g(\brr) \exp(-i \bk_0 \cdot \brr_0 ) =  \frac{1  }{(2 \pi)^{3/2}}
\int a(\bk_0 + \bq) \exp( i\bq \cdot \brr ) \di^3 q,
\end{eqnarray}
which, according to Eqs. (\ref{fo10}-\ref{fo15}), implies
\begin{eqnarray}\label{b300}
{a(\bk_0 + \bq) = G(\bq) \exp({-i \bk_0 \cdot \brr_0 }). }
\end{eqnarray}
were  $G(\bq)$ denotes the Fourier transform of $g(\brr)$. Now, to
complete the calculation we have to remember that we have defined
$g(\brr) = f(D \brr - D\brr_0)$ and that, according to Eq.
(\ref{a90}) we can write
\begin{eqnarray}\label{b310}
G(\bq) = F(D \bq) \exp(- i \bq \cdot \brr_0),
\end{eqnarray}
where Eqs. (\ref{fo40},\ref{b120}) have been used. By substituting
Eq. (\ref{b310}) into Eq. (\ref{b300}) we obtain
\begin{eqnarray}\label{b320}
\boxed{ a(\bk_0 + \bq) =  F(D \bq) \exp\left[-i (\bk_0 + \bq) \cdot \brr_0
\right]. }
\end{eqnarray}
This equation can be now used in Eq. (\ref{b260}) to obtain our
final result
\begin{eqnarray}\nonumber
\Psi_0(\brr, t) & = & \frac{e^{i(\bk_0 \cdot \brr - \omega_0 t)
}}{(2 \pi)^{3/2}} \int a(\bk_0 + \bq) \exp[ i\bq \cdot \brr - i
\delta \omega(\bk_0, \bq) t] \di^3 q \\ \nonumber
& = & \frac{e^{i(\bk_0 \cdot \brr - \omega_0 t) }}{(2 \pi)^{3/2}}
\int F(D \bq) e^{-i (\bk_0 + \bq) \cdot \brr_0 } \exp[ i\bq \cdot
\brr - i \delta \omega(\bk_0, \bq) t] \di^3 q \\  \nonumber
& = & \frac{e^{i[\bk_0 \cdot (\brr - \brr_0) - \omega_0 t] }}{(2
\pi)^{3/2}} \int F(D \bq)  \exp[ i\bq \cdot (\brr - \brr_0) - i
\delta \omega(\bk_0, \bq) t] \di^3 q \\ \label{b330}
& = & \frac{e^{i[\bk_0 \cdot (\brr - \brr_0) - \omega_0 t] }}{(2
\pi)^{3/2}} \int F(\bp)  e^{i[ ( D^T \bp )\cdot (\brr - \brr_0) -
\delta \omega(\bk_0, D^T \bp) t]} \di^3 p ,
\end{eqnarray}
where in the last line we made the change of variables
\begin{eqnarray}\label{b340}
\bq \rightarrow \bp \equiv D \bq.
\end{eqnarray}
From Eqs. (\ref{b100}), (\ref{b105}), and (\ref{b270}), it readily
follows that
\begin{eqnarray}\label{b350}
(D^T \bp)\cdot (\brr - \brr_0) = \bp\cdot D(\brr -\brr_0) = \bp
\cdot \bx',
\end{eqnarray}
and
\begin{eqnarray} \nonumber
\delta \omega(\bk_0, D^T \bp)
&  = &  \omega_0 \left[ 1 + 2 \vkappa_0 \cdot \frac{D^T \bp}{k_0}
+ \left( \frac{D^T \bp}{k_0}\right)^{2} \right]^{1/2} - \omega_0
\\
\label{b360}
 & = &
\omega_0 \left[ \left( 1 + 2 \frac{p_3}{k_0} + \frac{p^2}{k_0^2}
\right)^{1/2} - 1 \right],
\end{eqnarray}
where $(D^T \bp)^2 = (D^T \bp) \cdot (D^T \bp) = \bp D D^T \bp =
\bp \cdot \bp \equiv p^2$, and
\begin{eqnarray}\label{b370}
\vkappa_0 \cdot (D^T\bp) = (D \vkappa_0) \cdot \bp = D_{ij}
(\be_3')_j p_i = D_{ij}D^T_{j3} p_i = p_3.
\end{eqnarray}
Since the Dirac delta function in Eq. (\ref{b170}) implies that $2
p_3/k_0  = (p_3^2 -  p^2)/k_0^2$, it follows that we can rewrite
$\delta \omega(D^T \bp)$ as
\begin{eqnarray} \nonumber
\delta \omega(\bk_0, D^T \bp)
& = &
\omega_0 \left[ \left( 1 + 2 \frac{p_3}{k_0} + \frac{p^2}{k_0^2}
\right)^{1/2} - 1 \right] \\ \nonumber
&  = & \omega_0 \left[ \left( 1 + \frac{p_3^2}{k_0^2}
\right)^{1/2} - 1 \right] \\ \nonumber
&  = & \omega_0 \left[ \sqrt{ 1 + \left(\frac{p_1^2 + p_2^2}{2
k_0^2} \right)^2} - 1 \right]
 \\ \label{b380}
&  \simeq &  \frac{\omega_0}{2} \left(\frac{p_1^2 + p_2^2}{2 k_0^2}
\right)^2 + \ldots,
\end{eqnarray}
where  the last approximate equality holds in the paraxial limit
$|\bp|/{k_0} \ll 1$. Thus, we see that  for an input Gaussian
beam, the non-paraxial corrections to the carrying frequency
$\omega_0$, begin at fourth order. Finally, by using Eq. (\ref{b350}), we can rewrite Eq.
(\ref{b330}) as
\begin{eqnarray}\label{b390}
\Psi_0(\brr, t) = \frac{e^{i[\bk_0 \cdot (\brr - \brr_0) -
\omega_0 t] }}{(2 \pi)^{3/2}} \int F(\bq)  \exp{[ i \bq \cdot \bx'
 - i \tau(\bq) t]} \di^3 q ,
\end{eqnarray}
where we have defined $ \tau(\bq) \equiv \delta \omega(\bk_0,D^T
\bq)$. It is gratifying to see that after this tour de force we
re-obtained our starting result (\ref{b235}):
\begin{eqnarray}\nonumber
\Psi_0(\brr, t = 0) & =  & \frac{e^{i\bk_0 \cdot (\brr - \brr_0)
}}{(2 \pi)^{3/2}} \int F(\bq)  \exp{( i \bq \cdot \bx' )} \di^3 q
\\ \nonumber
& =  & e^{i\bk_0 \cdot (\brr - \brr_0)} f(\bx'=D \brr - D \brr_0)  \\
\nonumber
& \equiv  & e^{i\bk_0 \cdot (\brr - \brr_0)} g( \brr)\\
\label{b400}
& =  & \Psi(\brr, t = 0).
\end{eqnarray}
\subsection{Geometric reflected scalar beams}
Now that we learned how to represent a scalar Gaussian beam in an
arbitrary Cartesian frame, we can study how such representation
changes under reflection. In Sec. I, we showed that a generic
scalar field $\phi(\brr)$ transforms to  $\widetilde{\phi}(\brr) =
\phi(\widetilde{\brr})$ under geometric reflection with respect to the plane
of equation $z=0$. Therefore, by using Eq. (\ref{b390}) we can
write at once $\Psi_0(\brr, t) \rightarrow
\widetilde{\Psi}_0(\brr, t)$, where
\begin{eqnarray}\label{b410}
\widetilde{\Psi}_0(\brr, t) = \frac{e^{i[\bk_0 \cdot
(\widetilde{\brr} - \brr_0) - \omega_0 t] }}{(2 \pi)^{3/2}} \int
F(\bq)  \exp{[ i \bq \cdot (D \widetilde{\brr} - D \brr_0)
 - i \tau(\bq) t]} \di^3 q.
\end{eqnarray}
However, in this form, this equation is of little usefulness. In
order to gain insights, first we  {\textbf{a)}}  define a ``mirror image''
local frame $K''$ specular to $K'$ (as shown in Fig. 3), and then {\textbf{b})} we see how a scalar
field represented by $\phi(x,y,z)$ in $K$, and by
$\phi'(x',y',z')$ in $K'$, can be represented by
$\phi''(x'',y'',z'')$ in $K''$, where $\phi''(x'',y'',z'')$ is to
be found. The Cartesian frames definitions are illustrated in Figs. C1 and C2.

Thus, let $K'' \equiv (O''x''y''z'')$ be a Cartesian frame centered in
$\widetilde{\brr}_0 = -r_0 \widetilde{\vkappa}_0$, and let
 $\{\ve_1'', \ve_2'', \ve_3'' \} = \{\ve_1(\widetilde{\vkappa}_0), \ve_2(\widetilde{\vkappa}_0), \ve_3(\widetilde{\vkappa}_0) \}$
be three unit vectors pointing along the positive directions of the
axes $x''$, $y''$, and $z''$, respectively. They are defined via Eqs.
(\ref{a71}-\ref{a91}) with the substitution $\bk
\rightarrow \widetilde{\vkappa}_0$. As before, let $P$ denotes an
arbitrary point of coordinates $\overrightarrow{OP} \equiv \brr
\doteq (x_1,x_2,x_3) \equiv \brr$ in $K$, and coordinates
$\overrightarrow{O''P} \equiv \brr'' \doteq (x''_1,x''_2,x''_3)
\equiv \bx''$ in $K''$. The coordinates of the origin $O''$ of
$K''$, with respect to  $K$, are  equal to $\overrightarrow{OO''}
\equiv \widetilde{\brr}_0 \doteq
(\widetilde{x}_{01},\widetilde{x}_{02},\widetilde{x}_{03}) =
(x_{01},x_{02},-x_{03}) \equiv \widetilde{\brr}_0$ where, by definition,
 $\brr_0 = (x_{01},x_{02},x_{03})$.
\begin{flushright}
\begin{figure}[!hr]
\includegraphics[angle=0,width=9truecm]{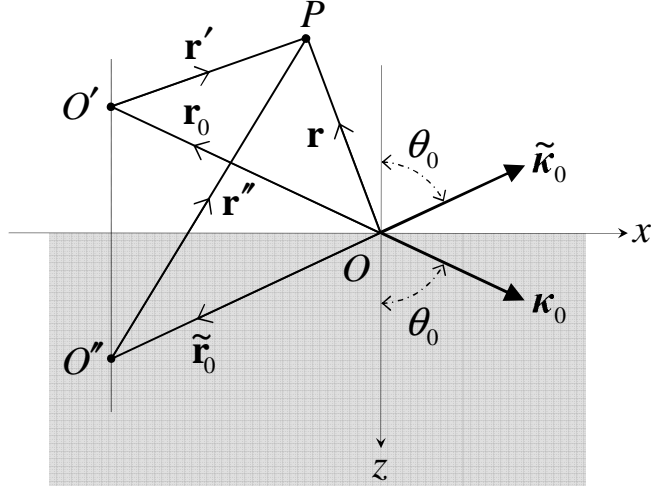}
\caption{\label{fig:4} Illustrating the  definitions of some vectors used in the text. }
\end{figure}
\end{flushright}
Then, the following equality it is trivially true
\begin{eqnarray}\label{b420}
\overrightarrow{OO''} + \overrightarrow{O''P} =
\overrightarrow{OP} \quad \Leftrightarrow \quad \widetilde{\brr}_0
+ \brr'' = \brr.
\end{eqnarray}
Therefore, we can always write
%
\begin{eqnarray}\nonumber
x_i'' & \equiv & \be_i'' \cdot \brr''
 \\  \label{b430}
& = & \be_i'' \cdot (\brr - \widetilde{\brr}_0)
 \\ \nonumber
& = & \sum_{j=1}^3 \be_i'' \cdot \be_j (x_j - \widetilde{x}_{0j})
 \\ \label{b440}
& \equiv & \sum_{j=1}^3 \widetilde{D}_{ij} (x_j -
\widetilde{x}_{0j}),
\end{eqnarray}
%
or, in vector notation
\begin{eqnarray}\label{b445}
\bx'' = \widetilde{D}(\brr  -\widetilde{\brr}_0).
\end{eqnarray}
 where we have defined
$\widetilde{D}_{ij} \equiv \be_i'' \cdot \be_j$.
It is easy to prove that $\widetilde{D}$ is orthogonal, that is
$\widetilde{D} \widetilde{D}^T = I$. Now, without loss of
generality we choose the Cartesian frame $(Oxyz)$ in such a way
that both $\vkappa_0$ and $\widetilde{\vkappa}_0$ lie in the $xz$
plane (the plane of incidence), that is we write
%
\begin{eqnarray}\label{b450}
\vkappa_0 & = & \be_1 \sin \theta_0  + \be_3 \cos \theta_0, \\
\label{b460}
 \widetilde{\vkappa}_0 & = & \be_1 \sin \theta_0  -
\be_3 \cos \theta_0.
\end{eqnarray}
%
%
In such a frame it is straightforward to calculate:
%
%
\begin{align}\label{b470}
D & = \left(%
\begin{array}{ccc}
  \cos \theta_0 & 0 & -\sin \theta_0 \\
  0 & 1 & 0 \\
  \sin \theta_0 & 0 & \cos \theta_0 \\
\end{array}%
\right), \\ \nonumber \\ \label{b480} \widetilde{D} & = \left(%
\begin{array}{ccc}
  -\cos \theta_0 & 0 & -\sin \theta_0 \\
  0 & 1 & 0 \\
  \sin \theta_0 & 0 & -\cos \theta_0 \\
\end{array}%
\right) = X D Z,
\end{align}
%
where we have defined the three ``reflection'' matrices $X, Y, Z$
as
\begin{eqnarray}\label{b490}
X \equiv \left(%
\begin{array}{ccc}
  -1 & 0 & 0 \\
  0 & 1 & 0 \\
  0 & 0 & 1 \\
\end{array}%
\right) , \;\; Y \equiv \left(%
\begin{array}{ccc}
  1 & 0 & 0 \\
  0 & -1 & 0 \\
  0 & 0 & 1 \\
\end{array}%
\right), \;\; Z \equiv \left(%
\begin{array}{ccc}
  1 & 0 & 0 \\
  0 & 1 & 0 \\
  0 & 0 & -1 \\
\end{array}%
\right).
\end{eqnarray}
Since $X^2 = Y^2 = Z^2 =I$, from Eq. (\ref{b480}) it readily
follows that $D = X \widetilde{D} Z$. Then, given any vector $\bu
\doteq (u_1,u_2,u_3)$ in $K$, one can use $Z$ to write
\begin{eqnarray}\label{b500}
\widetilde{\bu} = Z \bu,
\end{eqnarray}
from which it trivially follows that
%
%
\begin{eqnarray}\label{b510}
D \widetilde{\bu} = D Z \bu = X^2 D Z \bu = X \widetilde{D} \bu,\\
\label{b520}
 D \bu =  X^2 D Z^2 \bu = X^2 D Z \widetilde{\bu}  = X
\widetilde{D} \widetilde{\bu} .
\end{eqnarray}
%
%
Note that although we have derived the relations
(\ref{b480},\ref{b510},\ref{b520}) starting from the definitions
(\ref{b450}-\ref{b460}), they are independent from the latter. At
this point we can go back to Eq. (\ref{b410}) and notice that we
can rewrite
\begin{eqnarray}\label{b530}
\bk_0 \cdot (\widetilde{\brr} - \brr_0) =  \bk_0 \cdot Z (\brr -
\widetilde{ \brr}_0) = Z  \bk_0 \cdot (\brr - \widetilde{ \brr}_0)
=  \widetilde{\bk}_0 \cdot (\brr - \widetilde{ \brr}_0),
\end{eqnarray}
and
\begin{eqnarray}\label{b540}
D \widetilde{\brr} - D \brr_0 = X \widetilde{D} \brr - X
\widetilde{D} \widetilde{\brr}_0 = X \widetilde{D} (\brr -
\widetilde{\brr}_0) = X \bx'',
\end{eqnarray}
where Eq. (\ref{b445}) has been used. If we substitute Eq.
(\ref{b530})-(\ref{b540}) into Eq. (\ref{b410}), we finally obtain
\begin{eqnarray}\nonumber
\widetilde{\Psi}_0(\brr, t) &  = & \frac{e^{i[\widetilde{\bk}_0
\cdot (\brr - \widetilde{ \brr}_0) - \omega_0 t] }}{(2 \pi)^{3/2}}
\int F(\bq)  \exp{[ i \bq \cdot  X \widetilde{D} (\brr -
\widetilde{\brr}_0)
 - i \tau(\bq) t]} \di^3 q \\ \label{b550}
&  = & \frac{e^{i[\widetilde{\bk}_0 \cdot (\brr - \widetilde{
\brr}_0) - \omega_0 t] }}{(2 \pi)^{3/2}} \int F(\bq)  \exp{[ i \bq
\cdot  X \bx''
 - i \tau(\bq) t]} \di^3 q.
\end{eqnarray}
It should be noticed that at $t=0$ such expression reduces to
\begin{eqnarray}\label{b560}
\widetilde{\Psi}_0(\brr, t=0) = e^{i k_0 z'' } f(X \bx''),
\end{eqnarray}
where we used Eq. (\ref{b430}) to write
\begin{eqnarray}\label{b570}
\widetilde{\bk}_0 \cdot (\brr - \widetilde{ \brr}_0) = k_0
\vkappa_0 \cdot (\brr - \widetilde{ \brr}_0)= k_0 \ve_3'' \cdot
(\brr - \widetilde{ \brr}_0)=k_0 x_3'' \equiv k_0 z''.
\end{eqnarray}
Equation (\ref{b560}) is exactly what we expect for a Gaussian beam
centered at $\widetilde{\brr}_0$ and directed along
$\widetilde{\bk}_0$. It is also important to notice the argument $X
\bx''$ of the function $f$ that correctly account for the parity
inversion between local frames in reflection. In other words, if:
$\phi(x,y,z)$ and  $\widetilde{\phi}(x,y,z) =\phi(x,y,-z)$ are a
field and its mirror image as seen from the frame
$K$;  $\phi'(x',y',z')$ is the field $\phi(x,y,z)$ as seen from the
local frame $K'$, and, finally, $\phi''(x'',y'',z'')$ is the field
$\widetilde{\phi}(x,y,z)$ as seen from the local frame $K''$; then
$\phi''(\xi,\eta,\zeta)=\phi'(-\xi,\eta,\zeta)$, where $(
\xi,\eta,\zeta )$ is a set of three given real numbers:
\begin{eqnarray}\label{b580}
\begin{CD}
\phi(\brr) @ =\widetilde{ \phi}(Z \brr)\\
@VV{K'}V @VV{K''}V \\
\phi'(\bx') @ =  \phi''(X \bx')
\end{CD}
\end{eqnarray}
where $\bx' = D(\brr - \brr_0)$ and  $\bx'' = \widetilde{D}(\brr -
\widetilde{\brr}_0)$.
\section{Vector beams}
Up to now, we considered only scalar beams of the form
\begin{eqnarray}\label{c10}
\Psi(\brr ,t) = \frac{1}{(2 \pi)^{3/2}} \int a(\bk) \exp[i(\bk
\cdot \brr - \omega t)] \di^3 k,
\end{eqnarray}
which could be a faithful representation of an unpolarized beam.
However, if one is interested to polarization properties of the
field, such representation does not work and one has to pass to a
vector description. Since a polarized beam can be physically
obtained on a laboratory bench by sending an unpolarized beam
across a polarizer oriented, for example, at $\vf$, one could
naively think that it would be possible to promote the scalar
field $\Psi(\brr ,t)$ to the rank of a vector field ${\bm
\Psi}(\brr ,t)$, just by multiplying $\Psi(\brr ,t)$ by $\vf$:
\begin{eqnarray}\label{c20}
\Psi(\brr ,t) \rightarrow   \vPsi(\brr ,t) = \vf \Psi(\brr ,t),
\end{eqnarray}
where $\vf$ is a real- or complex-valued unit vector. This
equation is obviously wrong since the so-obtained field
$\vPsi(\brr ,t)$ is \emph{not} transverse:
\begin{eqnarray} \nonumber
\mathrm{div} \, \vPsi (\brr ,t) & = & \nabla \cdot \vPsi (\brr ,t)
\\ \nonumber
&  = & \sum_{i=1}^3 f_i \frac{\partial }{\partial x_i}\Psi(\brr
,t)
\\ \nonumber
&  = &\frac{i}{(2 \pi)^{3/2}} \int a(\bk)(\bk \cdot \vf )
\exp[i(\bk \cdot \brr - \omega t)] \di^3 k
\\ \label{c30}
& \neq 0 & \quad \mathrm{(in \; general).}
\end{eqnarray}
However, for any wave vector $\bk$ we can always write $\vf$ as
\begin{eqnarray}\nonumber
\vf &  = &  \left(\frac{\bk \bk}{k^2} \right) \vf + \left( I  -
\frac{\bk \bk}{k^2} \right) \vf \\ \nonumber
 &  = &  \mathcal{L}(\bk) \vf + \mathcal{T}(\bk) \vf \\ \label{c40}
& \equiv & \bff_\parallel (\bk) + \bff_\perp (\bk),
\end{eqnarray}
where $\bff_\perp(\bk)$ is genuinely transverse, namely $\bk \cdot
\bff_\perp(\bk) =0$, and Eq. (\ref{a90}) has been used. This fact suggests us the possibility to
substitute each scalar amplitude $a(\bk)$ in the plane-wave
expansion (\ref{c10}), with the vector amplitude $a(\bk)
\bff_\perp (\bk)$:
\begin{eqnarray}\label{c50}
a(\bk) \rightarrow   a(\bk) \bff_\perp (\bk),
\end{eqnarray}
thus obtaining a bona fide transverse field
\begin{eqnarray}\label{c60}
\vPsi(\brr ,t) = \frac{1}{(2 \pi)^{3/2}} \int a(\bk) \bff_\perp
(\bk) \exp[i(\bk \cdot \brr - \omega t)] \di^3 k.
\end{eqnarray}
Let $\{\ve_1(\bk), \ve_2(\bk), \ve_3(\bk) \}$ be an orthogonal and complete basis built following the recipe given in Appendix A.  By using such a vector basis, it is easy to see that we can rewrite  Eq. (\ref{c60}) as
\begin{eqnarray}\label{c140}
\vPsi(\brr ,t) = \frac{1}{(2 \pi)^{3/2}} \sum_{\lambda=1}^2 \int
\be_{\lambda}(\bk) a_\lambda(\bk)  \exp[i(\bk \cdot \brr - \omega
t)] \di^3 k,
\end{eqnarray}
where we have defined
\begin{eqnarray}\label{c150}
a_\lambda(\bk) \equiv a(\bk) \be_{\lambda}(\bk) \cdot \vf.
\end{eqnarray}
In conclusion, if we have a beam passing through a polarizer
oriented at $\vf$, it can be described by a vector field
$\vPsi(\brr ,t)$ such that
\begin{eqnarray}\label{c160}
\vPsi(\brr ,t) = \mathcal{M}(\brr,t) \vf,
\end{eqnarray}
where $\mathcal{M}(\brr,t) $ is a $3 \times 3$ matrix defined as
\begin{eqnarray}\label{c170}
\mathcal{M}(\brr,t) = \frac{1}{(2 \pi)^{3/2}} \int
 \mathcal{T}(\bk) a(\bk)  \exp[i(\bk \cdot \brr - \omega
t)] \di^3 k.
\end{eqnarray}

Now, we apply the previous formulas to the actual case of a
quasi-monochromatic Gaussian beam propagating along the positive
direction of the $z$-axis, whose equation is
\begin{eqnarray}\label{c180}
\psi(\brr,t) = \exp[i k_0(z - c t)]f(\brr),
\end{eqnarray}
where
\begin{eqnarray}\label{c190}
f(\brr) = \frac{- i L}{z - i L} \exp \left[ i \frac{k_0}{2} \left(
\frac{x^2 + y^2}{z - i L}  \right) \right].
\end{eqnarray}
Here, $\lambda_0 = 2 \pi /k_0$ is the carrying wavelength of the
beam, $c$ is the speed of light in vacuum, and $L = k_0 w_0^2 /2$ is
the Raleigh range of the beam whose waist at $z = 0$ is equal to:
$w(z = 0) = w_0$. In this Appendix we are not interested to the
temporal evolution of the beam, so we assume $t=0$ in all subsequent
formulas. Then, we rewrite Eq. (\ref{c10}) as
\begin{eqnarray}\nonumber
\Psi(\brr ,0) & = & \frac{e^{i k_0 z}}{(2 \pi)^{3/2}} \int a(\bk)
\exp[i \brr \cdot (\bk  - k_0 \be_z  )] \di^3 k \\ \label{c200}
& = & \frac{e^{i k_0 z}}{(2 \pi)^{3/2}} \int a(\bq + k_0 \be_z)
\exp(i \brr \cdot \bq ) \di^3 q,
\end{eqnarray}
where in the second line we have changed the variables of
integration from $\bk$ to $\bq = \bk  - k_0 \be_z $. If we impose
the condition $\Psi(\brr ,0) = \psi(\brr ,0)$, then from Eq.
(\ref{c180}) and Eq. (\ref{c200}), it readily follows that
\begin{eqnarray}\label{c210}
f(\brr) = \frac{1}{(2 \pi)^{3/2}}\int a(\bq + k_0 \be_z) \exp(i \brr
\cdot \bq ) \di^3 q,
\end{eqnarray}
which implies that $a(\bq + k_0 \be_z)$ coincides with the Fourier
transform $F(\bq)$ of $f(\brr)$: $a(\bq + k_0 \be_z) = F(\bq)$,
namely
\begin{eqnarray}\label{c220}
a(\bq + k_0 \be_z) = \frac{(2 \pi)^{1/2} L}{k_0} \exp \left[
-\frac{L}{2 k_0} (q_1^2 + q_2^2) \right] \delta \left( q_3 +
\frac{q_1^2 + q_2^2 }{2 k_0} \right),
\end{eqnarray}
where   Eq. (\ref{b170}) has been used. Note that the presence of
the Dirac delta in the Fourier transform $F(\bq)$ of $f(\brr)$ is
not accidental. In fact,  given a function $f(\brr)$ solution of the
paraxial wave equation:
%
\begin{eqnarray}\label{c222}
\left( \frac{\partial^2}{\partial x^2} + \frac{\partial^2}{\partial
y^2} + 2 i k_0 \frac{\partial}{\partial z} \right)f(\brr) = 0,
\end{eqnarray}
it is always possible to write
\begin{eqnarray}\nonumber
0 & = & \left( \frac{\partial^2}{\partial x^2} +
\frac{\partial^2}{\partial
y^2} + 2 i k_0 \frac{\partial}{\partial z} \right)f(\brr)\\
\nonumber
& = & \left( \frac{\partial^2}{\partial x^2} +
\frac{\partial^2}{\partial y^2} + 2 i k_0 \frac{\partial}{\partial
z} \right) \left[ \frac{1}{(2 \pi)^{3/2}}\int F(\bq ) \exp(i \brr
\cdot \bq ) \di^3 q \right]\\
\label{c223}
& = & \frac{- 2 k_0}{(2 \pi)^{3/2}}  \int F(\bq ) \left( q_3 +
\frac{q_1^2 + q_2^2}{2 k_0} \right)\exp(i \brr \cdot \bq ) \di^3 q.
\end{eqnarray}
This equation implies that
\begin{eqnarray}\label{c224}
F(\bq ) \left( q_3 + \frac{q_1^2 + q_2^2}{2 k_0} \right) =  0 \quad
\forall (q_1,q_2,q_3) \in \mathbb{R}^3,
\end{eqnarray}
namely $F(\bq)$ must be zero outside the $2D$ surface of  equation $
q_3 + \frac{q_1^2 + q_2^2}{2 k_0} = 0$. However, in order to have
$f(\brr) \neq 0$, it is clear that $F(\bq )$ must be infinite
\emph{upon} such surface. Thus, it must be always possible to write
\begin{eqnarray}\label{c225}
F(\bq ) = \mathcal{F}(\bq ) \; \delta \left( q_3 + \frac{q_1^2 +
q_2^2 }{2 k_0} \right),
\end{eqnarray}
where $\mathcal{F}(\bq )$ is finite everywhere in $\mathbb{R}^3$.
 At this point it is a
trivial task to rewrite Eq. (\ref{c170}) as
\begin{eqnarray}\nonumber
\mathcal{M}(\brr ,0) & = & \frac{e^{i k_0 z}}{(2 \pi)^{3/2}} \int
\mathcal{T} (\bq + k_0 \be_z) a(\bq + k_0 \be_z) \exp(i \brr \cdot
\bq) \di^3 q \\  \nonumber
& = & \frac{e^{i k_0 z  }L}{2 \pi k_0} \int \di q_1 \di q_2 \Biggl\{
\exp \left[ (i x)q_1 + (i y) q_2  -\frac{L}{2 k_0} (q_1^2 + q_2^2)
\right]  \Biggr.
\\ \nonumber
& &   \times \Biggl. \int \mathcal{T} (q_1,q_2, q_3 + k_0) \exp
\left( i z q_3  \right) \delta \left( q_3 + \frac{q_1^2 + q_2^2 }{2
k_0} \right) \di q_3 \Biggr\} \\ \nonumber
%
%
& = & \frac{e^{i k_0 z  }L}{2 \pi k_0} \int   \exp \left[ (i x)q_1 +
(i y) q_2  -\frac{L + i z}{2 k_0} (q_1^2 + q_2^2) \right]
\\ \label{c230}
& &   \times \Biggl. \mathcal{T} \left( q_1, q_2 ,   k_0 - {(q_1^2 +
q_2^2 )}/({2 k_0})  \right)  \di q_1 \di q_2,
\end{eqnarray}
where the symmetric transverse tensor $\mathcal{T} (\bq + k_0
\be_z)$, after integration with respect to the variable $q_3$, takes
the value
\begin{eqnarray}\nonumber
\mathcal{T} \bigl( q_1, q_2 , k_0 - {(q_1^2 + q_2^2
)}/({2 k_0}) \bigr)
& = & \left(
\begin{array}{ccc}
            1 - \frac{Q_1^2}{1 +Q_\perp^4/4} &
            - \frac{Q_1 Q_2}{1 +Q_\perp^4/4} &
            - \frac{Q_1(1 -Q_\perp^2/2 )}{1 +Q_\perp^4/4} \\
            - \frac{Q_1 Q_2}{1 +Q_\perp^4/4} &
            1 - \frac{Q_2^2}{1 +Q_\perp^4/4} &
            - \frac{Q_2(1 -Q_\perp^2/2 )}{1 +Q_\perp^4/4} \\
            - \frac{Q_1(1 -Q_\perp^2/2 )}{1 +Q_\perp^4/4}  &
            - \frac{Q_1(1 -Q_\perp^2/2 )}{1 +Q_\perp^4/4}
            & \frac{Q_\perp^2}{1 +Q_\perp^4/4} \\
\end{array}
\right)
\\ \label{c240}
 & \cong  & \left(
\begin{array}{ccc}
            1 -  Q_1^2 & - Q_1 Q_2 & - Q_1 \\
            -  Q_1 Q_2 &  1 - Q_2^2 & - Q_2 \\
           - Q_1 & - Q_2 & Q_1^2 + Q_2^2 \\
\end{array}
\right).
\end{eqnarray}
In the first line of the equation above, we have defined the
dimensionless variables $\{ Q_\lambda \equiv q_\lambda / k_0
\}_{\lambda = 1}^2$, and $Q_\perp^2 \equiv Q_1^2 + Q_2^2$; while in
the second line we have made a Taylor expansion about $Q_1 = 0 =
Q_2$ up to the second order, which is justified for paraxial beams.

To illustrate in detail the formalism just developed, let us apply
it to calculate the vector structure of a Gaussian beam passing
trough a polarizer oriented along the $x$-axis: $\vf = \ve_x$. From
Eq. (\ref{c160}) it follows that
\begin{eqnarray}\nonumber
{\bm \Psi}(\brr,0) & = & \mathcal{M}(\brr,0)\ve_x\\ \nonumber
& = & \sum_{i=1}^3 \ve_i \ve_i \cdot \left[ \mathcal{M}(\brr,0)\ve_x  \right]\\
\label{c242}
& = & \mathcal{M}_{11}(\brr,0)\ve_x + \mathcal{M}_{21}(\brr,0)\ve_y
+\mathcal{M}_{31}(\brr,0)\ve_z
\end{eqnarray}
The matrix elements $\mathcal{M}_{ij}$ can be easily calculated in
the second order approximation by using Eqs. (\ref{c230}-\ref{c240})
as follows:
\begin{eqnarray}\nonumber
\mathcal{M}_{11}(\brr ,0) & \cong & \frac{e^{  (i k_0 z ) }k_0 L}{2
\pi} \int  \exp \left[ (i k_0 x)Q_1 + (i k_0 y) Q_2 \right] \left( 1
- Q_1^2  \right)
\\ \nonumber
& &   \times \Biggl. \exp \left\{ -\frac{1}{2} \left[ k_0 L + (i k_0
z) \right] (Q_1^2 + Q_2^2) \right\} \di Q_1 \di Q_2
\\ \nonumber
& = & \frac{e^{  X_3 } \Lambda}{2 \pi} \int \exp \left( X_1 Q_1 +
X_2 Q_2 \right) \left( 1 -  Q_1^2  \right)
\\ \nonumber
& &   \times \Biggl. \exp \left[ -\frac{1}{2} \left( \Lambda + X_3
\right) (Q_1^2 + Q_2^2) \right] \di Q_1 \di Q_2
\\ \nonumber
& = & \frac{e^{  X_3 } \Lambda}{2 \pi} \int \left( 1 -  Q_1^2
\right)
\\  \label{c250}
& &   \times \Biggl. \exp \Biggl[ -\frac{1}{2} \sum_{ \alpha, \beta
=1 }^2 Q_\alpha (A^{-1})_{\alpha \beta} Q_\beta + \sum_{ \alpha =1
}^2 Q_\alpha X_\alpha \Biggr]  \di^2 Q,
\end{eqnarray}
where we have defined the dimensionless variables
\begin{eqnarray}\label{c260}
X_1 \equiv i k_0 x , \quad X_2 \equiv i k_0 y  , \quad X_3 \equiv i
k_0 z ,  \quad \Lambda \equiv k_0 L,
\end{eqnarray}
and the covariance matrix $A$ of elements
\begin{eqnarray}\label{c270}
A_{\alpha \beta} \equiv \frac{1}{\left( \Lambda + X_3 \right)}
\delta_{\alpha \beta}, \qquad (\alpha, \beta \in \{ 1,2 \}).
\end{eqnarray}
Gaussian integrals are easy to calculate and the following result
can be found in any graduate textbook:
\begin{eqnarray}\nonumber
G &  = &  \int  \exp \left[ -\frac{1}{2} \sum_{ \alpha, \beta =1 }^2
Q_\alpha (A^{-1})_{\alpha \beta} Q_\beta + \sum_{ \alpha =1 }^2
Q_\alpha X_\alpha \right]  \di^2 Q\\ \label{c280}
&  = &  2 \pi ( \det A)^{1/2} \exp \left[ \frac{1}{2} \sum_{ \alpha,
\beta =1 }^2 X_\alpha A_{\alpha \beta} X_\beta \right] .
\end{eqnarray}
In our case, with $A$ given by Eq. (\ref{c270}), we obtain
\begin{eqnarray}\label{c285}
G  =  \frac{ 2 \pi }{\Lambda + X_3} \exp \Biggl[ \frac{1}{2} \left(
\frac{X_1^2 + X_2^2}{\Lambda + X_3} \right) \Biggr].
\end{eqnarray}
By using this result we can easily calculate the following two
quantities that we need:
\begin{eqnarray}\nonumber
G_\alpha &  = &  \int Q_\alpha \exp \Biggl[ -\frac{1}{2} \sum_{
\alpha, \beta =1 }^2 Q_\alpha (A^{-1})_{\alpha \beta} Q_\beta +
\sum_{ \alpha =1 }^2 Q_\alpha X_\alpha \Biggr]  \di^2 Q\\
\nonumber
&  = &  \frac{\partial G}{\partial X_\alpha}\\
\label{c290}
&  = & G \frac{ X_\alpha }{\Lambda + X_3},
\end{eqnarray}
and
\begin{eqnarray}\nonumber
G_{\alpha \beta} &  = &  \int Q_\alpha Q_\beta \exp \Biggl[
-\frac{1}{2} \sum_{ \alpha, \beta =1 }^2 Q_\alpha (A^{-1})_{\alpha
\beta} Q_\beta +
\sum_{ \alpha=1 }^2 Q_\alpha X_\alpha \Biggr]  \di^2 Q\\
\nonumber
&  = &  \frac{\partial^2 G}{\partial X_\alpha \partial X_\beta }\\
\label{c300}
&  = & G \left[ \frac{ X_\alpha X_\beta}{(\Lambda + X_3)^2} +
 \frac{ \delta_{\alpha \beta}}{\Lambda + X_3}
  \right].
\end{eqnarray}
Now we have all the ingredients we need to calculate the three
matrix elements required to evaluate Eq. (\ref{c242}). In sequence,
first we calculate $\mathcal{M}_{11}(\brr ,0)$:
\begin{eqnarray}\nonumber
\mathcal{M}_{11}(\brr ,0)
& = & \frac{e^{  X_3 } \Lambda}{2 \pi}  \left( G - \frac{\partial^2
G}{\partial X_1^2 } \right) + \ldots
\\ \nonumber
& = &  \frac{- i L}{z - i L} \exp \left[ i k_0 z + i \frac{k_0}{2}
\left( \frac{x^2 + y^2}{z - i L}  \right) \right]
\\ \nonumber
& &   \times \Biggl[ 1 + \frac{i}{k_0} \frac{1}{z - i L} - \left(
\frac{x}{z - i L} \right)^2 \Biggr] + \ldots \\ \label{c310}
&=&\psi(\brr,0)\Biggl[ 1 + \frac{i}{k_0} \frac{1}{z - i L} - \left(
\frac{x}{z - i L} \right)^2 \Biggr] + \ldots,
\end{eqnarray}
where Eq. (\ref{c180}-\ref{c190}) have been used. Then we calculate
$\mathcal{M}_{21}(\brr ,0)$:
\begin{eqnarray}\nonumber
\mathcal{M}_{21}(\brr ,0)
& = &  \frac{e^{  X_3 } \Lambda}{2 \pi} \int \exp \left( X_1 Q_1 +
X_2 Q_2 \right) \left(  -  Q_1 Q_2 \right)
\\ \nonumber
& &   \times \Biggl. \exp \left\{ -\frac{1}{2} \left( \Lambda + X_3
\right) (Q_1^2 + Q_2^2) \right\} \di Q_1 \di Q_2
\\ \nonumber
& = & - \frac{e^{  X_3 } \Lambda}{2 \pi} \frac{\partial^2
G}{\partial X_2 \partial X_1 }\\ \nonumber
& = & \frac{- i L}{z - i L} \exp \left[ i k_0 z + i \frac{k_0}{2}
\left( \frac{x^2 + y^2}{z - i L}  \right) \right] \left[ - \frac{x
y}{( z- i L)^2}  + \ldots\right]
\\ \label{c320}
& = & \psi(\brr,0)\left[ - \frac{x y}{( z- i L)^2}  + \ldots\right],
\end{eqnarray}
and, finally, we calculate $\mathcal{M}_{31}(\brr ,0)$:
\begin{eqnarray}\nonumber
\mathcal{M}_{31}(\brr ,0)
& = &  \frac{e^{  X_3 } \Lambda}{2 \pi} \int \exp \left( X_1 Q_1 +
X_2 Q_2 \right) \left(  -  Q_1  \right)
\\ \nonumber
& &   \times \Biggl. \exp \left\{ -\frac{1}{2} \left( \Lambda + X_3
\right) (Q_1^2 + Q_2^2) \right\} \di Q_1 \di Q_2
\\ \nonumber
& = & - \frac{e^{  X_3 } \Lambda}{2 \pi} \frac{\partial G}{\partial
X_1 }
\\ \nonumber
& = & \frac{- i L}{z - i L} \exp \left[ i k_0 z + i \frac{k_0}{2}
\left( \frac{x^2 + y^2}{z - i L}  \right) \right] \left[ \frac{x }{
z- i L}  + \ldots\right]
\\ \label{c330}
& = & \psi(\brr,0)\left[ \frac{x }{ z- i L}  + \ldots\right].
\end{eqnarray}
By collecting all the results above we can obtain the final
expression for a unpolarized Gaussian beam crossing a polarizer
oriented at $\vf = \ve_x$, exact up to the second order terms:
\begin{eqnarray}\label{c340}
{ \bm \Psi(\brr ,0) } \cong \psi(\brr,0) \left\{ \be_x \Biggl[ 1 + \frac{i}{k_0} \frac{1}{z - i L} - \left(
\frac{x}{z - i L} \right)^2 \Biggr] - \be_y \frac{xy
}{( z- i L)^2} + \be_z \frac{x }{ z- i L} \right\} .
\end{eqnarray}
As expected, the dominant zero-order term is the ``naive-guess''
vector field $\psi(\brr,0) \be_x$. The first order correction gives
a longitudinal contribution $\propto \ve_z$ to the field. Finally,
the so-called ``crossed-polarization'' term $\propto \ve_y$ amounts
to a second order correction. A plot of the intensities of the three
terms is reported in Figs. 5-7.  These results should be compared with the similar ones presented in ref. \cite{EandS}.
\begin{flushright}
\begin{figure}[]
\includegraphics[angle=0,width=7truecm]{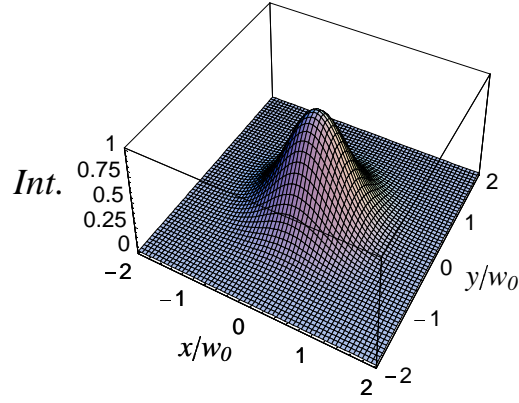}
\caption{\label{fig:6} Transverse intensity profile due to  the  first term $\propto \be_x$ (dominant polarization term) in Eq. (\ref{c340}).}
\end{figure}
\begin{figure}[]
\includegraphics[angle=0,width=7truecm]{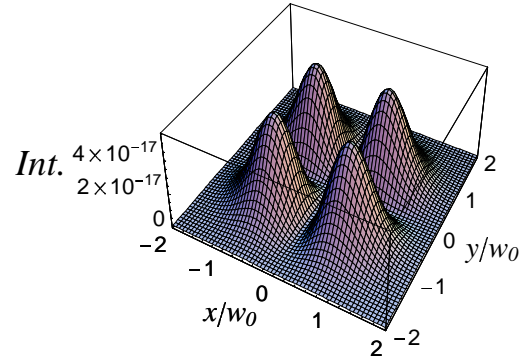}
\caption{\label{fig:7} Transverse intensity profile due to  the second  term $\propto \be_y$ (cross polarization term) in Eq. (\ref{c340}). }
\end{figure}
\begin{figure}[]
\includegraphics[angle=0,width=7truecm]{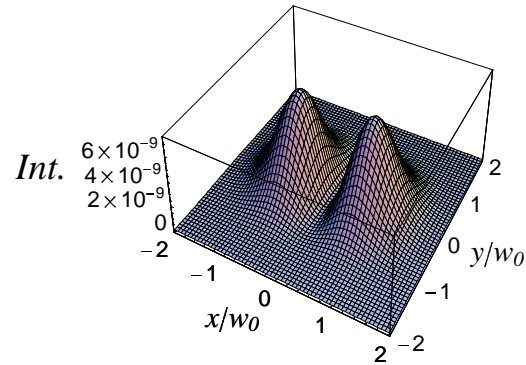}
\caption{\label{fig:8} Transverse intensity profile due to  the third  term $\propto \be_z$ (longitudinal polarization term) in Eq. (\ref{c340}). }
\end{figure}
\end{flushright}
\section{Expansion of the polarization vectors}
Let $\{ \ve_1(\bk), \ve_2(\bk),\ve_3(\bk) \}$ be an orthogonal and complete basis for $\mathbb{R}^3$ defined as in Appendix A, where $\bk \in \mathbb{R}^3$ is a given vector. Each basis element $\ve_i(\bk)$ can be expanded in a Taylor series around an arbitrary vector $\bk = \bk_0$. If $\bk_0$ denotes the ``carrying'' wave vector of a well collimated light beam, then it is reasonable to keep only the first and second order terms in the Taylor expansion, and write
\begin{align}\nonumber
 \ve_i(\bk_0 + \bq) & \simeq  \ve_i(\bk_0) +  q_l \left.\frac{\partial  \ve_i(\bk)}{\partial  k_l} \right|_{\bk = \bk_0} + \frac{1}{2}   q_l q_m\left. \frac{\partial^2  \ve_i(\bk)}{\partial  k_l \partial  k_m} \right|_{\bk = \bk_0} + \dots \\ \label{f10}
 & \equiv  \bee_i^{(0)} +  \frac{1}{k_0}q_l \mathbf{e}_i^{(l)} + \frac{1}{2 k_0^2}   q_l q_m \mathbf{e}_i^{(lm)} + \dots ,
\end{align}
where we have defined
\begin{align}\label{f15}
\bee_i^{(0)} & \equiv  \bee_i^{(0)}(\bk_0)  =  \ve_i(\bk_0),
\\ \label{f16}
\bee_i^{(l)} & \equiv  \bee_i^{(l)}(\bk_0)  =  k_0 \left.   \frac{\partial  \ve_i(\bk)}{\partial  k_l} \right|_{\bk = \bk_0},
\\ \label{f17}
\bee_i^{(lm)} & \equiv  \bee_i^{(lm)}(\bk_0)  =   k_0^2 \left.  \frac{\partial^2  \ve_i(\bk)}{\partial  k_l \partial  k_m} \right|_{\bk = \bk_0}.
\end{align}
Since we have always the freedom to choose the right-handed basis $\{ \ve_x, \ve_y, \ve_z \} \equiv \{ \ve_1, \ve_2, \ve_3 \}$ that defines orientation of the Cartesian frame $(Oxyz)$, we can take $\bk_0$ laying on the $xz$-plane:
\begin{equation}\label{f20}
\bk_0 = k_0 \left( \ve_x \sin \theta_0 + \ve_z \cos \theta_0  \right)= k_0 \begin{pmatrix}
               \sin \theta_0 \\
               0 \\
               \cos \theta_0 \\
             \end{pmatrix},
\end{equation}
where $k_0 \equiv |\bk_0|$. In this basis is a straightforward task to calculate, by using the definitions given in Eqs. (\ref{f15}-\ref{f17}), the three zero-order vectors $\{ \bee_i^{(0)}\}_{i=1}^3$, the nine first order vectors $\{ \bee_i^{(l)} \}_{i,l=1}^3$, and the twenty-seven second-order vectors $\{ \bee_i^{(lm)} \}_{i,l,m=1}^3$. Explicitly, the zero-order vectors are
\begin{equation}\label{f40}
\bee_1^{(0)} = \begin{pmatrix}
               \cos \theta_0 \\
               0 \\
               -\sin \theta_0 \\
             \end{pmatrix}, \qquad
\bee_2^{(0)} = \begin{pmatrix}
               0 \\
               1 \\
               0 \\
             \end{pmatrix}, \qquad
\bee_3^{(0)} = \begin{pmatrix}
               \sin \theta_0 \\
               0 \\
               \cos \theta_0 \\
             \end{pmatrix}.
\end{equation}
The first-order vectors are
\begin{equation}\label{f50}
\bee_1^{(1)} = \begin{pmatrix}
               -\cos \theta_0 \sin \theta_0 \\
               0 \\
               -\cos^2 \theta_0 \\
             \end{pmatrix}, \qquad
\bee_2^{(1)} = \begin{pmatrix}
               0 \\
               0 \\
               0 \\
             \end{pmatrix}, \qquad
\bee_3^{(1)} = \begin{pmatrix}
               \cos^2 \theta_0 \\
               0 \\
               -\cos \theta_0 \sin \theta_0 \\
             \end{pmatrix},
\end{equation}
\begin{equation}\label{f60}
\bee_1^{(2)} = \begin{pmatrix}
               0 \\
               \cot \theta_0 \\
               0 \\
             \end{pmatrix}, \qquad
\bee_2^{(2)} = \begin{pmatrix}
               -\csc \theta_0 \\
               0 \\
               0 \\
             \end{pmatrix}, \qquad
\bee_3^{(2)} = \begin{pmatrix}
               0 \\
               1 \\
               0 \\
             \end{pmatrix},
\end{equation}
\begin{equation}\label{f70}
\bee_1^{(3)} = \begin{pmatrix}
               \sin^2 \theta_0 \\
               0 \\
               \cos \theta_0 \sin \theta_0 \\
             \end{pmatrix}, \qquad
\bee_2^{(3)} = \begin{pmatrix}
               0 \\
               0 \\
               0 \\
             \end{pmatrix}, \qquad
\bee_3^{(3)} = \begin{pmatrix}
               -\cos \theta_0 \sin \theta_0 \\
               0 \\
               \sin^2 \theta_0 \\
             \end{pmatrix}.
\end{equation}
Since $ \mathbf{e}_i^{(lm)} =  \mathbf{e}_i^{(ml)}$, the only relevant second-order vectors are
\begin{equation}\label{f80}
\bee_1^{(11)} = \begin{pmatrix}
               -\frac{1}{4} \left( \cos \theta_0 + 3 \cos 3 \theta_0 \right) \\
               0 \\
               3 \cos^2 \theta_0 \sin \theta_0 \\
             \end{pmatrix}, \quad
\bee_2^{(11)} = \begin{pmatrix}
               0 \\
               0 \\
               0 \\
             \end{pmatrix}, \quad
\bee_3^{(11)} = \begin{pmatrix}
               -3 \cos^2 \theta_0 \sin \theta_0 \\
               0 \\
               -\frac{1}{4} \left( \cos \theta_0 + 3 \cos 3 \theta_0 \right)\\
             \end{pmatrix},
\end{equation}
\begin{equation}\label{f90}
\bee_1^{(12)} = \begin{pmatrix}
               0 \\
               -\cos \theta_0 - \cot \theta_0 \csc \theta_0 \\
               0 \\
             \end{pmatrix}, \quad
\bee_2^{(12)} = \begin{pmatrix}
               \csc^2 \theta_0 \\
               0 \\
               0 \\
             \end{pmatrix}, \quad
\bee_3^{(12)} = \begin{pmatrix}
               0 \\
               -\sin \theta_0 \\
              0\\
             \end{pmatrix},
\end{equation}
\begin{equation}\label{f100}
\bee_1^{(13)} = \begin{pmatrix}
               \frac{1}{2} \left(1 + 3 \cos 2 \theta_0 \right)\sin \theta_0 \\
               0 \\
               \frac{1}{4} \left(\cos \theta_0 + 3 \sin 3 \theta_0  \right) \\
             \end{pmatrix}, \quad
\bee_2^{(13)} = \begin{pmatrix}
               0 \\
               0 \\
               0 \\
             \end{pmatrix}, \quad
\bee_3^{(13)} = \begin{pmatrix}
               -\frac{1}{4} \left(\cos \theta_0 + 3 \cos 3 \theta_0 \right) \\
               0 \\
               -\frac{1}{4} \left(\sin \theta_0 - 3 \sin 3 \theta_0  \right) \\
             \end{pmatrix},
\end{equation}
\begin{equation}\label{f110}
\bee_1^{(22)} = \begin{pmatrix}
               -\cos \theta_0 - \cot \theta_0 \csc \theta_0 \\
               0 \\
               - \cos \theta_0 \cot \theta_0 \\
             \end{pmatrix}, \quad
\bee_2^{(22)} = \begin{pmatrix}
               0 \\
               -\csc^2 \theta_0 \\
               0 \\
             \end{pmatrix}, \quad
\bee_3^{(22)} = \begin{pmatrix}
               -\sin \theta_0  \\
               0 \\
               -\cos \theta_0 \\
             \end{pmatrix},
\end{equation}
\begin{equation}\label{f120}
\bee_1^{(23)} = \begin{pmatrix}
               0 \\
               \sin \theta_0 \\
               0 \\
             \end{pmatrix}, \quad
\bee_2^{(23)} = \begin{pmatrix}
               0 \\
               0 \\
               0 \\
             \end{pmatrix}, \quad
\bee_3^{(23)} = \begin{pmatrix}
               0 \\
               -\cos \theta_0 \\
               0 \\
             \end{pmatrix},
\end{equation}
\begin{equation}\label{f130}
\bee_1^{(33)} = \begin{pmatrix}
               -3 \cos \theta_0 \sin^2 \theta_0 \\
               0 \\
               \frac{1}{4} \left( \sin \theta_0 - 3 \sin 3 \theta_0 \right) \\
             \end{pmatrix}, \quad
\bee_2^{(33)} = \begin{pmatrix}
               0 \\
               0 \\
               0 \\
             \end{pmatrix}, \quad
\bee_3^{(33)} = \begin{pmatrix}
               -\frac{1}{4} \left( \sin \theta_0 - 3 \sin 3 \theta_0 \right) \\
               0 \\
               -3 \cos \theta_0 \sin^2 \theta_0 \\
             \end{pmatrix}.
\end{equation}

From the equations above it is very simple to obtain the corresponding formulas for the expansions of the polarization vectors defined around the ``reflected'' vector $\widetilde{\bk}_0$, by noting that
\begin{align}\label{f140}
\bee_1^{(\widetilde{0})} & \equiv  \bee_1^{(0)}(\widetilde{\bk}_0)  =  -\left. \bee_1^{({0})} \right|_{\theta_0 \rightarrow - \theta_0},
\\ \label{f141}
\bee_2^{(\widetilde{0})} & \equiv  \bee_2^{(0)}(\widetilde{\bk}_0)  =  \bee_2^{({0})} ,
\\ \label{f142}
\bee_3^{(\widetilde{0})} & \equiv  \bee_3^{(0)}(\widetilde{\bk}_0) =   -\left. \bee_3^{({0})} \right|_{\theta_0 \rightarrow - \theta_0},
\end{align}
and
%
\begin{align}\label{f150}
\bee_1^{(\widetilde{l})} & \equiv  \bee_1^{(l)}(\widetilde{\bk}_0)  =  \left. \bee_1^{(l)} \right|_{\theta_0 \rightarrow - \theta_0},
\\ \label{f151}
\bee_2^{(\widetilde{l})} & \equiv  \bee_2^{(l)}(\widetilde{\bk}_0)  =  \bee_2^{(l)},
\\ \label{f152}
\bee_3^{(\widetilde{l})} & \equiv   \bee_3^{(l)}(\widetilde{\bk}_0)  =\left. \bee_3^{(l)} \right|_{\theta_0 \rightarrow - \theta_0},
\end{align}
and
%
\begin{align}\label{f160}
\bee_1^{(\widetilde{lm})} & \equiv  \bee_1^{(lm)}(\widetilde{\bk}_0)  = - \left. \bee_1^{(lm)} \right|_{\theta_0 \rightarrow - \theta_0},
\\ \label{f161}
\bee_2^{(\widetilde{lm})} & \equiv  \bee_2^{(lm)}(\widetilde{\bk}_0)  =  \bee_2^{(lm)},
\\ \label{f162}
\bee_3^{(\widetilde{lm})} & \equiv   \bee_3^{(lm)}(\widetilde{\bk}_0)  = - \left. \bee_3^{(lm)} \right|_{\theta_0 \rightarrow - \theta_0}.
\end{align}

A quantity that often appears during calculations, is the diadic product
\begin{equation}\label{f170}
\ve_\mu(\widetilde{\bk}_0 + \widetilde{\bq})\ve_\nu({\bk_0} + {\bq}), \qquad(\mu,\nu \in \{ 1,2\}),
\end{equation}
which can be expressed in terms of the expansions (\ref{f10}), up to second order terms, as
\begin{align}\nonumber
\ve_\mu(\widetilde{\bk}_0 + \widetilde{\bq})\ve_\nu({\bk_0} + {\bq}) \simeq & \; \bee_\mu^{(\widetilde{0})} \bee_\nu^{(0)} + \frac{q_i}{k_0} \Bigl[
 \bee_\mu^{(\widetilde{0})} \bee_\nu^{(i)}  + (-1)^{\delta_{i3}}\bee_\mu^{(\widetilde{i})} \bee_\nu^{(0)}
 \Bigr] \\ \nonumber
 & + \frac{1}{2 k_0^2} q_i q_j \Bigl[\ \bee_\mu^{(\widetilde{0})} \bee_\nu^{(i j)}  + 2 (-1)^{\delta_{i3}} \bee_\mu^{(\widetilde{i})} \bee_\nu^{(j)} +  (-1)^{\delta_{i3}+\delta_{j3}} \bee_\mu^{(\widetilde{i j})} \bee_\nu^{(0)}\Bigr]  \\ \label{f182}
\equiv & \; \mathbf{E}_{\mu \nu} + Q_i \mathbf{E}^{(i)}_{\mu \nu} + \frac{1}{2} Q_i Q_j \mathbf{E}^{(i j)}_{\mu \nu},
\end{align}
were we have introduced the dimensionless variables $\{ Q_i \equiv q_i / k_0 \}_{i=1}^3$. The $4 + 12 +36$  matrices (each $3 \times 3$) $\mathbf{E}_{\mu \nu}, \, \mathbf{E}^{(i)}_{\mu \nu}$ and $\mathbf{E}^{(ij)}_{\mu \nu}$, can be calculated straightforwardly from the following definitions:
\begin{align}\label{f190}
\mathbf{E}_{\mu \nu} \equiv & \;  \bee_\mu^{(\widetilde{0})} \bee_\nu^{(0)}, \\ \label{f200}
\mathbf{E}_{\mu \nu}^{(i)} \equiv & \;  \bee_\mu^{(\widetilde{0})} \bee_\nu^{(i)}  + (-1)^{\delta_{i3}}\bee_\mu^{(\widetilde{i})} \bee_\nu^{(0)}, \\ \label{f210}
\mathbf{E}^{(i j)}_{\mu \nu} \equiv & \; \bee_\mu^{(\widetilde{0})} \bee_\nu^{(i j)}  + 2 (-1)^{\delta_{i3}} \bee_\mu^{(\widetilde{i})} \bee_\nu^{(j)} +  (-1)^{\delta_{i3}+\delta_{j3}} \bee_\mu^{(\widetilde{i j})} \bee_\nu^{(0)}.
\end{align}

Note that the presence of terms like $\cot \theta_0$ and $\csc \theta_0$ make the expansion above not utilizable for ``small'' values of $\theta_0$. More precisely, it is possible to show that the expansion (\ref{f10}) breaks down when
\begin{equation}\label{f220}
\sin \theta_0 \leq \sin \theta \equiv \frac{|\bk_0 \times (\bk_0 + \bq)|}{k_0|\bk_0 + \bq|},
\end{equation}
where Eq. (2.9) of ref. \cite{DandG} has been used.
\section{Paraxial formulation}
In this appendix we consider an alternative and simpler formulation for the study of the reflection of a Maxwell-Gaussian beam by a planar surface. Basically, we use here the formalism developed by Haus and Pan in Ref. \cite{HausandPan}. The geometry of the problem and the notation we use are those ones fixed in appendix C. Let $\mathbf{A}^\inp(x',y',z',t)$ the vector potential of the impinging electromagnetic field, as seen from the local frame $K'$. By hypothesis, it obeys the Lorentz gauge condition. We assume that $\mathbf{A}^\inp(x',y',z',t)$ can be written in the paraxial approximation, as
\begin{equation}\label{g10}
\mathbf{A}^\inp(x',y',z',t) = \left[ \be_1' \psi_1^\inp (\bx') + \be_2' \psi_2^\inp (\bx') \right] \exp(- i \omega_0 t)
\end{equation}
where, from Eq. (C6), $\bx' = D(\brr - \brr_0)$, and
\begin{equation}\label{g20}
\psi_\alpha^\inp (\bx') = u_\alpha f(x',y',z') \exp(i k_0 z'), \qquad (\alpha = 1,2),
\end{equation}
where $f(x',y',z')$ is defined by Eq. (C9), and $u_\alpha \in \mathbb{C}: |u_1|^2 +  |u_2|^2 =1$. Following appendix C, we define $g(\brr) = f(\bx' = D \brr - D \brr_0)$, and  use Eqs. (C15,C16,C18) to rewrite
\begin{align}\nonumber
\psi_\alpha^\inp (\bx' = D \brr - D \brr_0) & = u_\alpha \Psi(\brr,0) \\ \nonumber
 & = u_\alpha \Psi_0(\brr,0)\\ \label{g30}
 & = u_\alpha  \frac{e^{i\bk_0 \cdot \brr
}}{(2 \pi)^{3/2}} \int a(\bk_0 + \bq) \exp( i\bq \cdot \brr) \di^3 q,
\end{align}
where Eq. (C28) has been used.

Let $\mathbf{A}^\out(x'',y'',z'',t)$ the vector potential of the reflected electromagnetic field, as seen from the local frame $K''$:
\begin{equation}\label{g40}
\mathbf{A}^\out(x'',y'',z'',t) = \left[ \be_1'' \psi_1^\out (\bx'') + \be_2'' \psi_2^\out (\bx'') \right] \exp(- i \omega_0 t).
\end{equation}
Since reflection causes $\brr \rightarrow \widetilde{\brr}$ and $a(\bk_0 + \bq) \rightarrow r_\alpha(\bk_0 + \bq)a(\bk_0 + \bq)$, we can write at once
\begin{align}\label{g50}
\psi_\alpha^\out (\bx'' = \widetilde{D}\brr - \widetilde{D}\widetilde{\brr}) & = u_\alpha  \frac{e^{i\bk_0 \cdot \widetilde{\brr}
}}{(2 \pi)^{3/2}} \int r_\alpha(\bk_0 + \bq)a(\bk_0 + \bq) \exp( i\bq \cdot \widetilde{\brr}) \di^3 q,
\end{align}
where there is \emph{no} summation over repeated Greek indices.
At this point, it is easy to follow the procedure outlined in Sec. III in order to obtain
\begin{align}\nonumber
\psi_\alpha^\out (\bx'') & = u_\alpha  \frac{e^{ik_0 z''
}}{(2 \pi)^{3/2}} \int r_\alpha(\bk_0 + D^T \bq)F(\bq) \exp( i\bq \cdot X \bx'') \di^3 q\\ \nonumber
 & =  e^{ik_0 z''} \left[ u_\alpha  \frac{k_0L}{2 \pi} r_\alpha(\bk_0) G_\alpha(\bx'') \right]\\ \label{g60}
 & \equiv  e^{ik_0 z''} a_\alpha(\bx''),
\end{align}
where $G_\alpha(\bx'')$ is defined by Eq. (53). From Eqs. (8-9) of Ref. \cite{HausandPan} and the equation above, it is easy to calculate the time-independent parts of the reflected electric and magnetic fields:
\begin{align}\label{g70}
\mathbf{E} (\bx'') & = i \omega_0  e^{i k_0 z''} \sum_{\alpha =1}^2 \Bigl[
 \be_\alpha'' a_\alpha + \be_3''\frac{i}{k_0} \left( \be_\alpha'' \cdot \nabla_T'' a_\alpha \right) \Bigr], \\ \label{g80}
\mathbf{B} (\bx'') & = i k_0  e^{i k_0 z''} \sum_{\alpha =1}^2 \Bigl[
 (\be_3'' \times \be_\alpha'') a_\alpha + \frac{i}{k_0} \left( \be_\alpha'' \times \nabla_T'' a_\alpha \right) \Bigr],
\end{align}
where
\begin{equation}\label{g90}
\nabla_T'' a_\alpha =\be_1'' \frac{\partial}{\partial x''}a_\alpha(\bx'') + \be_2'' \frac{\partial}{\partial y''}a_\alpha(\bx'').
\end{equation}
These expressions are both analytical and easy to handle. They can be used for the final step, namely the computation of the time average Poynting vector $\mathbf{S} (\bx'')$:
\begin{equation}\label{g100}
\mathbf{S} (\bx'') = \frac{1}{2 \mu_0} \mathrm{Re}\left[\mathbf{E} (\bx'') \times \mathbf{B}^* (\bx'') \right] .
\end{equation}
The flux of the Poynting vector through the detector's surface gives the measured intensity. Assuming that the normal to the detector's surface is directed parallel to $\be_3''$, we have to calculate the position of the center of the function $S_3 (\bx'')$ at the position $z'' = Z_D$ of the detector:
\begin{equation}\label{g110}
S_3 (x'',y'',Z_D) = \frac{\omega_0 k_0}{2 \mu_0} \left( |a_1(x'',y'',Z_D)|^2 + |a_2(x'',y'',Z_D)|^2 \right).
\end{equation}
Thus, the GH and the IF shifts, as measured in $K''$, are simply:
\begin{equation}\label{g120}
L_{GH} = \frac{\displaystyle{\sum_{\alpha =1}^2\int x |a_\alpha(x,y,Z_D)|^2   \di x \di y}}{\displaystyle{\sum_{\alpha =1}^2\int |a_\alpha(x,y,Z_D)|^2 \di x \di y}},
\end{equation}
\begin{equation}\label{g130}
L_{IF} = \frac{\displaystyle{ \sum_{\alpha =1}^2\int y|a_\alpha(x,y,Z_D)|^2   \di x \di y}}{\displaystyle{\sum_{\alpha =1}^2 \int |a_\alpha(x,y,Z_D)|^2 \di x \di y}},
\end{equation}
respectively. Note that these integrals are all Gaussian, so that they can be calculated analytically.

It is quite clear that this approximate formulation can give a correct expression for the GH shift, but not for the IF shift.
\section{Polarization of light beams}
\subsection*{Notation}
The \emph{scalar} product between two complex-valued three-dimensional vectors $\vec{u}$ and $\vec{v}$ is defined as
\begin{equation}\label{n10}
 \left( \vec{u}, \vec{v}  \right) =  \sum_{i=1}^3 u_i^* v_i = u_i^* v_i = \left( \vec{v}, \vec{u}  \right)^* \in \mathbb{C},
\end{equation}
where summation over repeated indices is understood, and $u_i, v_i$ are the Cartesian component of $\vec{u}$ and $\vec{v}$, respectively, with respect to an arbitrary orthogonal reference frame. Unit vectors are denoted by the `` \verb"hat" '' symbol:
\begin{equation}\label{n20}
\hat{u} = \frac{\vec{u}}{\sqrt{(\vec{u},\vec{u})}} \equiv \frac{\vec{u}}{u}, \qquad (\hat{u},\hat{u})=1,
\end{equation}
where $u \equiv \sqrt{(\vec{u},\vec{u})}$. The diadic product $ \vec{u} \, \vec{v}^{\, \dagger}$  represents a $3 \times 3$ matrix whose elements are defined as
\begin{equation}\label{n25}
( \vec{u} \, \vec{v}^{\, \dagger})_{ij} = u_i v^*_j.
\end{equation}
Operators are denoted by capital letters either as   $\mathbb{P}$ or  $\mathcal{P}$;  their matrix representations in a given basis are always written in ``bold'' characters as  $\mathbf{P}$.
\subsection*{Arbitrary elliptical polarizers}
Let us consider the action of a polarizer upon a quasi-monochromatic well collimated beam of light that crosses it non-orthogonally. Let $(Oxyz)$ be the laboratory frame specified by the real-valued basis
\begin{equation}\label{n27}
\{\hat{x},\hat{y},\hat{z} \} \equiv \{\hat{e}_1,\hat{e}_2,\hat{e}_3 \}.
\end{equation}
An elliptical polarizer on the laboratory bench is characterized by its \emph{axis} $\hat{z}$ and its \emph{orientation} $\hat{p}$:
\begin{equation}\label{n30}
\hat{p} = p_x \hat{x} + p_y \hat{y}, 
\end{equation}
where $p_x, p_y$ are complex-valued numbers such that $|p_x|^2 +  |p_y|^2 =1$. This vector represents an optical device made by three optical elements: a quarter wave plate, a linear polarizer, and another quarter wave plate (see Ref. \cite{BickelandBailey}). This is easily seen by writing explicitly the Jones matrices $\mathbf{C}(\alpha)$ and $\mathbf{P}(\beta)$ of a compensator and a linear polarizer, respectively, in the two-dimensional laboratory basis $\{\hat{x}, \hat{y} \}$:
\begin{equation}\label{n32}
\mathbf{C}(\alpha) = \left(
                       \begin{array}{cc}
                         e^{i \alpha} & 0 \\
                         0 & e^{-i \alpha} \\
                       \end{array}
                     \right),\qquad
    \mathbf{P}(\beta) = \left(
                       \begin{array}{cc}
                         \cos^2 \beta & \sin \beta \cos \beta \\
                         \sin \beta \cos \beta & \sin^2 \beta \\
                       \end{array}
                     \right).
\end{equation}
These well known matrix representations hold when the impinging light beam propagates along the $z$-axis.  If, without loss of generality, we rewrite $p_x$ and $p_y$ as
\begin{equation}\label{n33}
 p_x =e^{- i \alpha} \cos \beta, \qquad  p_y =e^{ i \alpha} \sin \beta,
\end{equation}
then it is easy to check, via a straightforward calculation, the validity of the following relations:
\begin{align}\nonumber
\hat{p}\hat{p}^\dagger & = \left(
                                      \begin{array}{cc}
                                        p_x p_x^* &  p_x p_y^* \\
                                         p_y p_x^* &  p_y p_y^* \\
                                      \end{array}
                                    \right) \\ \nonumber
                                    &= \mathbf{C}(-\alpha)\mathbf{P}(\beta)\mathbf{C}(\alpha)\\ \label{n34}
                                    & = \left(
                                      \begin{array}{cc}
                                        \cos^2 \beta &  e^{-2 i \alpha} \sin \beta \cos \beta \\
                                         e^{2 i \alpha} \sin \beta \cos \beta  &  \sin^2 \beta \\
                                      \end{array}
                                    \right).
\end{align}

Now, let's turn back to the three-dimensional problem, and let $\hat{k}$ be a real-valued unit vector denoting the main direction of propagation of a quasi-monochromatic well collimated beam of light whose electric field vector can be written as
\begin{equation}\label{n35}
\vec{E}_0(\vec{r},t) = \left( E_{x'} \hat{x}' + E_{y'} \hat{y}'\right)e^{i\frac{\omega}{c}[(\hat{k},\vec{r}) -  c t]} \equiv \vec{E} e^{i\frac{\omega}{c}[(\hat{k},\vec{r}) -  c t]},
\end{equation}
where
\begin{equation}\label{n37} \{\hat{x}',\hat{y}',\hat{z}' \equiv \hat{k} \} \equiv \{\hat{e}_1',\hat{e}_2',\hat{e}_3' \},
\end{equation}
is a local Cartesian basis ``attached'' to the beam whose axis $\hat{z}'$ coincides with the beam direction of propagation $\hat{k}$. If with $\theta$ we denote the angles between $\hat{z}$ and $\hat{k}$, namely $(\hat{z},\hat{k}) = \cos \theta$, then we can chose the basis  $\{\hat{e}_1',\hat{e}_2',\hat{e}_3' \}$ as
\begin{align}\label{n38}
 \hat{e}_1'  &=  \hat{e}_1 \cos \theta  \cos \phi  + \hat{e}_2 \cos \theta
\sin \phi  - \hat{e}_3 \sin \theta,
 \\ \label{a61}  \hat{e}_2'
&=  -\hat{e}_1 \sin \phi  + \hat{e}_2 \cos  \phi ,
 \\ \label{a62}  \hat{e}_3'  &= \hat{e}_1 \sin \theta  \cos \phi  + \hat{e}_2 \sin \theta
\sin \phi  + \hat{e}_3 \cos \theta.
\end{align}
The electric field $\vec{E}$ is \emph{transverse}, that is $(\vec{E},\hat{k})$=0. In an equivalent manner, we can rewrite this condition as an eigenvalue equation
\begin{equation}\label{n39}
\mathcal{T}(\hat{k})\vec{E} = \vec{E},
\end{equation}
where $\mathcal{T}(\hat{k})$  is the operator that projects into the space orthogonal to the propagation direction $\hat{k}$. It is defined as
\begin{equation}\label{n40}
\mathcal{T}(\hat{k}) = \mathbb{I} - \hat{k}\hat{k}.
\end{equation}
The action of the polarizer upon the field $\vec{E}$ can be found by requiring that the polarization vector $\vec{\tau}$ of the transmitted beam lie in the complex plane defined by the polarizer orientation $\hat{p}$ and the propagation vector $\hat{k}$ of the beam. This condition is automatically satisfied by tacking the projection of $\hat{p}$ upon the space transverse to $\hat{k}$, namely
\begin{equation}\label{n50}
\vec{\tau} = \mathcal{T}(\hat{k}) \hat{p} = \hat{p} - \hat{k} ( \hat{k},\hat{p} ) \equiv \hat{p} - \alpha \hat{k}
\end{equation}
where $\alpha \equiv  ( \hat{k},\hat{p} )$ and $(\vec{\tau},\vec{\tau}) = 1  - |\alpha|^2 $. Note that $\vec{\tau}$ is a unit vector only when the beam impinges orthogonally to the polarizer, that is when $\alpha =0$. Thus, when the beam of light crosses the polarizer oriented at $\hat{p}$, its electric field vector becomes parallel to $\hat{\tau}$ according to
\begin{equation}\label{n52}
\vec{E} \rightarrow \vec{\tau} \bigl( \vec{\tau}, \vec{E} \bigr)  = \mathbb{T} \vec{E},
\end{equation}
where we have defined the operator projector $\mathbb{T}$ as
\begin{equation}\label{n55}
\mathbb{T} \equiv \vec{\tau} \, \vec{\tau}^{\; \dagger} =  \hat{p} \, \hat{p}^{\; \dagger} -
\bigl( \alpha \hat{k}\hat{p}^\dagger + \alpha^* \hat{p} \hat{k}\bigr) + |\alpha|^2 \hat{k} \hat{k}.
\end{equation}
It is easy to check that it satisfies the following properties
\begin{align}\label{n57}
\mathbb{T}\hat{k} & =\vec{0}, \qquad \mathrm{tr} \, \mathbb{T} =  (\vec{\tau},\vec{\tau}) = 1  - |\alpha|^2.
\\
\mathbb{T}^\dagger & = \mathbb{T}, \qquad \mathbb{T}^2 = \left( \mathrm{tr} \, \mathbb{T} \right) \mathbb{T}.
\end{align}
\subsection*{Input and output beams}
So, we have an input beam directed along the axis $\hat{k}$ whose electric field has Cartesian components $(E_{x'}, E_{y'})$ in the local ``beam'' frame $(O \,x'y'z')$. Moreover, the polarizer orientation vector $\hat{p}$ has components  $(p_{x}, p_{y})$ in the laboratory frame $(O \,xyz)$. Therefore, if we want to write Eq. (\ref{n52}) either in the beam or in the laboratory frame, we need to know how the electric field and polarizer orientation vector representations changes by passing from a frame to the other. Let us define the Cartesian components of the electric field $\{ E_i\}_{i=1}^3$ and $\{ E_i'\}_{i=1}^3$ in the laboratory and in the beam frames, respectively, by the following relations:
\begin{equation}\label{n90}
\vec{E} = \sum_{i=1}^3 \hat{e}_i (\hat{e}_i,\vec{E}) \equiv \sum_{i=1}^3 \hat{e}_i E_i,
\end{equation}
\begin{equation}\label{n100}
\vec{E} = \sum_{i=1}^3 \hat{e}_i' (\hat{e}_i',\vec{E}) \equiv \sum_{i=1}^3 \hat{e}_i' E_i'.
\end{equation}
The change of basis matrix $\mathbf{\Lambda}$ that permits to pass from one representation to the other is defined via its elements as
\begin{equation}\label{n110}
\Lambda_{ij} = (\hat{e}_i,\hat{e}_j'), \qquad \Lambda_{ij}^\dagger = (\hat{e}_i',\hat{e}_j),
\end{equation}
or, in explicit form,
\begin{equation}\label{n120}
{ \bm \Lambda} = \left(
                   \begin{array}{ccc}
                     \cos \theta  \cos \phi & -\sin \phi & \sin \theta  \cos \phi \\
                     \cos \theta
\sin \phi & \cos \phi & \sin \theta  \sin \phi \\
                     -\sin \theta & 0 & \cos \theta \\
                   \end{array}
                 \right),
\end{equation}
where ${ \bm \Lambda}{ \bm \Lambda}^\dagger ={ \bm \Lambda}^\dagger{ \bm \Lambda} = \mathbf{I}$.

By means of the electric field vector $\vec{E}$ we can describe any \emph{polarized} beam of light. However, if we need to describe partially polarized light, we must use the coherency operator $\mathbb{J}$ whose elements are defined in the laboratory and in the beam frames as
\begin{equation}\label{n130}
J_{ij} = \bigl(\hat{e}_i, \mathbb{J} \hat{e}_j \bigr) = \bigl\langle E_i E_j^*\bigr\rangle,
\end{equation}
\begin{equation}\label{n140}
J_{ij}' = \bigl(\hat{e}_i', \mathbb{J} \hat{e}_j' \bigr) = \bigl\langle E_i' {E_j'}^*\bigr\rangle,
\end{equation}
respectively, where angular brackets $\langle \cdots \rangle$ denotes either temporal or spatial (or both), averages. From a straightforward calculation it follows that
\begin{equation}\label{n150}
\mathbf{J}' =  { \bm \Lambda}^\dagger \mathbf{J} { \bm \Lambda},
\end{equation}
where, explicitly, for the input beam
\begin{equation}\label{n155}
\mathbf{J}_\mathrm{in}' =  \left(
                 \begin{array}{ccc}
                   \bigl\langle E_{x'} E_{x'}^*\bigr\rangle & \bigl\langle E_{x'} E_{y'}^*\bigr\rangle & 0 \\
                   \bigl\langle E_{y'} E_{x'}^*\bigr\rangle & \bigl\langle E_{y'} E_{y'}^*\bigr\rangle & 0 \\
                   0 & 0 & 0 \\
                 \end{array}
               \right)
.
\end{equation}

Now, following the same line of reasoning, we write the projection operator $\mathbb{T}$  in both the laboratory and in the beam frames,  as
\begin{equation}\label{n160}
T_{ij} = \left(\hat{e}_i, \mathbb{T} \hat{e}_j \right) = \left(\hat{e}_i, \vec{\tau} \, \vec{\tau}^{\; \dagger} \hat{e}_j \right) = \left(\hat{e}_i, \vec{\tau} \right) \left( \vec{\tau}, \hat{e}_j \right) \equiv t_i t_j^*,
\end{equation}
\begin{equation}\label{n170}
T_{ij}' = \left(\hat{e}_i', \mathbb{T} \hat{e}_j' \right) = \left(\hat{e}_i', \vec{\tau} \, \vec{\tau}^{\; \dagger} \hat{e}_j' \right) = \left(\hat{e}_i', \vec{\tau} \right) \left( \vec{\tau}, \hat{e}_j' \right) \equiv t_i' {t_j'}^*,
\end{equation}
respectively, where we have defined the representation vectors $\vec{t}, \vec{t}\,'$ as
\begin{equation}\label{n175}
\vec{\tau} = \sum_{i=1}^3\hat{e}_i t_i = \sum_{i=1}^3 \hat{e}_i' t_i'.
\end{equation}
It is easy to see that we can pass from the beam to the laboratory representation of the polarizer via the following relation:
\begin{equation}\label{n180}
\mathbf{T}' =  { \bm \Lambda}^\dagger \mathbf{T} { \bm \Lambda}.
\end{equation}

Now we have all the ingredients to write the relation between input and output beams in the frame we want. In particular, from a straightforward calculation it follows that if we write such relation as
\begin{equation}\label{n190}
\mathbb{J}_\mathrm{out} =  \mathbb{T} \mathbb{J}_\mathrm{in}  \mathbb{T}^\dagger,
\end{equation}
then, in the laboratory and in the beam frames we have
\begin{equation}\label{n200}
\mathbf{J}_\mathrm{out} =  \mathbf{T} \mathbf{J}_\mathrm{in}  {\mathbf{T}}^\dagger,
\end{equation}
\begin{equation}\label{n205}
\mathbf{J}_\mathrm{out}' =  \mathbf{T}' \mathbf{J}_\mathrm{in}'  {\mathbf{T}'}^\dagger,
\end{equation}
respectively.
\subsection*{Example}
As an application of the formalism developed above, let us consider the case of an unpolarized input light beam directed along $\hat{k}$, and represented in the beam frame by the following diagonal coherency matrix
\begin{equation}\label{n210}
\mathbf{J}_\mathrm{in}' = \frac{1}{2}\left(
                            \begin{array}{ccc}
                              1 & 0 & 0 \\
                              0 & 1 & 0 \\
                              0 & 0 & 0 \\
                            \end{array}
                          \right).
\end{equation}
This matrix is real-valued, so it does not contain any complex phase term. Moreover, let suppose to have a circular polarizer represented in the laboratory frame by the orientation vector
\begin{equation}\label{n220}
\hat{p} = \frac{1}{\sqrt{2}}\left( \hat{x} + i \hat{y} \right).
\end{equation}
By applying Eq. (\ref{n34}) we see immediately that in the laboratory $(O x y)$ frame
\begin{equation}\label{n225}
\hat{p}\hat{p}^\dagger = \frac{1}{2} \left(
                                       \begin{array}{cc}
                                         1 & -i \\
                                         i & 1 \\
                                       \end{array}
                                     \right) =  \mathbf{C}(-\pi/4)\mathbf{P}(\pi/4)\mathbf{C}(\pi/4).
\end{equation}
Now, from Eqs. (\ref{n50},\ref{n120},\ref{n175}) it is easy to calculate both $\vec{t}$ and $\vec{t}'$, obtaining
\begin{align}\nonumber
 \vec{t} & =  \frac{1}{\sqrt{2}}\left(
                                        \begin{array}{c}
                                          1 - e^{i \phi} \cos\phi \sin^2 \theta \\
                                          i - e^{i \phi} \sin\phi \sin^2 \theta\\
                                          - e^{i \phi} \sin\theta \cos \theta \\
                                        \end{array}
                                      \right) \\ \nonumber \\ \label{n230}
                                      & \cong  \frac{1}{\sqrt{2}}\left(
                                        \begin{array}{c}
                                          1 \\
                                          i \\
                                          0 \\
                                        \end{array}
                                      \right)+\theta \frac{e^{i \phi}}{\sqrt{2}}\left(
                                        \begin{array}{c}
                                          0 \\
                                          0 \\
                                          -1 \\
                                        \end{array}
                                      \right)  + \theta^2 \frac{e^{i \phi}}{\sqrt{2}}\left(
                                        \begin{array}{c}
                                          -\cos \phi \\
                                          -\sin \phi \\
                                          0 \\
                                        \end{array}
                                      \right) + \ldots,
\end{align}
\begin{align}\nonumber
 \vec{t}\,' & =  e^{i \phi} \frac{1}{\sqrt{2}}\left(
                                        \begin{array}{c}
                                          \cos\theta \\
                                          i \\
                                          0 \\
                                        \end{array}
                                      \right)
                                      \\ \nonumber \\\label{n240} & \cong  e^{i \phi} \left[\frac{1}{\sqrt{2}}\left(
                                        \begin{array}{c}
                                          1 \\
                                          i \\
                                          0 \\
                                        \end{array}
                                      \right) - \theta^2 \frac{1}{\sqrt{2}}\left(
                                        \begin{array}{c}
                                          1/2 \\
                                          0 \\
                                          0 \\
                                        \end{array}
                                      \right) + \ldots \right].
\end{align}
These equation are interesting for several reasons: First of all, apart from an irrelevant phase factor, at zero-order $\vec{t}$ and $\vec{t}'$ coincides with a circularly polarized unit vector. The phase factor present in the expression of $\vec{t}'$ is simply due to the non-univocal definition of the spherical coordinate $\phi$ when $\theta = 0$. Conventionally, in this case one puts $\phi =0$, thus removing the factor $ e^{i \phi} $. Second, while in the beam frame the corrections start from the second order term in the Taylor expansion of $ \vec{t}\,'$ around $\theta =0$, in the laboratory frame a first order (longitudinal) correction term arise. Third, for $\theta \neq 0$, the output beam polarization vector $\tau$ has no longer unit modulus:
\begin{equation}\label{n250}
(\vec{\tau}, \vec{\tau})=(\vec{t}, \vec{t})=(\vec{t}\,', \vec{t}\,') = \frac{1}{4}\left[3 + \cos (2 \theta) \right] \cong 1 - \frac{\theta^2}{2} + O(\theta)^4.
\end{equation}
Finally, by using Eqs. (\ref{n160}-\ref{n170},\ref{n200}-\ref{n205},\ref{n230}-\ref{n240}) we can write, after a straightforward calculation, the polarization state of the beam seen from both the laboratory and the beam frames, as
\begin{align}\label{n260}
\mathbf{J}_\mathrm{out} & \cong \frac{1}{4}\left(
                            \begin{array}{ccc}
                              1 & -i & 0 \\
                              i & 1 & 0 \\
                              0 & 0 & 0 \\
                            \end{array}
                          \right) - \frac{\theta}{4}\left(
                            \begin{array}{ccc}
                              0 & 0 & e^{- i \phi} \\
                              0 & 0 & i e^{- i \phi}\\
                               e^{ i \phi} & -ie^{ i \phi} & 0 \\
                            \end{array}
                          \right) \\ \nonumber
                          & + \frac{\theta^2}{4}\left(
                            \begin{array}{ccc}
                             -\frac{3}{2} - 2 \cos(2 \phi) & \frac{3}{2}i - 2 \sin(2 \phi) & 0 \\
                              -\frac{3}{2}i - 2 \sin(2 \phi) & -\frac{3}{2} + 2 \cos(2 \phi) & 0 \\
                              0 & 0 & 1 \\
                            \end{array}
                          \right) + \ldots
\end{align}
\begin{equation}\label{n270}
\mathbf{J}_\mathrm{out}' \cong \frac{1}{4}\left(
                            \begin{array}{ccc}
                              1 & -i & 0 \\
                              i & 1 & 0 \\
                              0 & 0 & 0 \\
                            \end{array}
                          \right) - \frac{\theta^2}{8}\left(
                            \begin{array}{ccc}
                              3 & -i & 0 \\
                              i & 1 & 0 \\
                              0 & 0 & 0 \\
                            \end{array}
                          \right) + \ldots
\end{equation}
As expected, both representations give a pure circularly polarized beam at zero order (remember that the input beam was unpolarized, so that the relative $\pi /2$ phase between field components was acquired by passing through the circular polarizer). Once again, the corrections are of the first order in the laboratory frame and second order in the beam frame.
To conclude, note that about half of the intensity of the input beam was  lost by passing across the polarizer:
\begin{equation}\label{n280}
\mathrm{tr}\, \mathbf{J}_\mathrm{in}' = 1 \rightarrow \mathrm{tr}\, \mathbf{J}_\mathrm{out}' = \frac{1}{32}[3 + \cos(2\theta)]^2 \cong \frac{1}{2} - \frac{\theta^2}{2} + \ldots
\end{equation}
%

%

\begin{thebibliography}{10}

\bibitem{BandWBook}
Max Born and Emil Wolf.
\newblock {\em Principles of optics}.
\newblock Cambridge University Press, Cambridge, UK, 7 edition, 2003.

\bibitem{KandS}
S.~Kozaki and H.~Sakurai.
\newblock Characteristic of a gaussian beam at a dielectric interface.
\newblock {\em J. Opt. Soc. Am.}, 68(4):508, April 1978.

\bibitem{Tamir}
T.~Tamir.
\newblock Nonspecular phenomena in beam fields reflected by multilayered media.
\newblock {\em J. Opt. Soc. Am. A}, 3(4):558, April 1986.

\bibitem{Nasalski}
Wojciech Nasalski.
\newblock Longitudinal and transverse effects of nonspecular reflection.
\newblock {\em J. Opt. Soc. Am. A}, 13(1):172, January 1996.

\bibitem{LandM}
Gary~D. Landry and Theresa~A. Maldonado.
\newblock Gaussian beam transmission and reflection from a general anisotropic
  multilayer structure.
\newblock {\em Appl. Opt.}, 35(30):5870, October 1996.

\bibitem{Barton}
John~P. Barton.
\newblock Electromagnetic field for a focused light sheet incident on a plane
  surface.
\newblock {\em J. Opt. Soc. Am. A}, 22(5):978, May 2005.

\bibitem{Gragg}
Robert~F. Gragg.
\newblock The total reflection of a compact wave group: Long range transmission
  in a waveguide.
\newblock {\em Am. J. Phys.}, 56(12):1092, December 1988.

\bibitem{CandQ}
K.~W. Chiu and J.~J. Quinn.
\newblock On the goos-h\"{a}nchen effect: A simple example of a time delay
  scattering process.
\newblock {\em Am. J. Phys.}, 40(12):1847, December 1972.

\bibitem{McGandC}
M.~McGuirk and C.~K. Carniglia.
\newblock An angular spectrum representation approach to the goos-h\"{a}nchen
  shift.
\newblock {\em J. Opt. Soc. Am.}, 67(1):103, January 1977.

\bibitem{LaiEtAl}
H.~M. Lai, C.~W. Kwok, Y.~W. Loo, and B.~Y. Xu.
\newblock Energy-flux pattern in the goos-h\"{a}nchen effect.
\newblock {\em Phys. Rev. E}, 62(5):7330, 2000.

\bibitem{LeungEtAl}
P.~T. Leung, C.~W. Chen, and H.-P. Chiang.
\newblock Large negative goos-h\"{a}nchen shift at metal surfaces.
\newblock {\em Opt. Commun.}, 276:206, 2007.

\bibitem{CandI}
O.~Costa de~Beauregard and C.~Imbert.
\newblock Quantized longitudinal and transverse shifts associated with total
  internal reflection.
\newblock {\em Phys. Rev. Lett.}, 28(18):1211, May 1972.

\bibitem{PillonEtAl}
Frank Pillon, Herv\'{e} Gilles, and Sylvain Girard.
\newblock Experimental observation of the imbert-fedorov transverse
  displacement after a single total reflection.
\newblock {\em Appl. Opt.}, 43(9):1863, Marchr 2004.

\bibitem{Li}
Chun-Fang Li.
\newblock Unified theory for goos-h\"{a}nchen and imbert-fedorov effects.
\newblock {\em Phys. Rev. A}, 76:013811, 2007.

\bibitem{EandS}
W.~L. Erikson and Surendra Singh.
\newblock Polarization properties of maxwell-gaussian laser beams.
\newblock {\em Phys. Rev. E}, 49(6):5778, June 1994.

\bibitem{EsquivelEtAl}
Raoul Esquivel, Carlos Villarreal, and W.~Louis Moch\'{a}n.
\newblock Exact surface impedance formulation of the casimir force: Application
  to spatially dispersive metals.
\newblock {\em Phys. Rev. A}, 68:052103, 2003.

\bibitem{Yao}
Carl Yao.
\newblock Magnetism and mirror symmetry.
\newblock {\em Am. J. Phys.}, 63(6):520, June 1995.

\bibitem{DandG}
Ivan~H. Deutsch and John~C. Garrison.
\newblock Paraxial quantum propagation.
\newblock {\em Phys. Rev. A}, 43(3):2498, March 1991.

\bibitem{Vandegrift}
Guy Vandegrift.
\newblock The diffraction and spreading of a wavepacket.
\newblock {\em Am. J. Phys.}, 72(3):404, March 2004.

\bibitem{Brownstein}
K.~R. Brownstein.
\newblock The whole-partial derivative.
\newblock {\em Am. J. Phys.}, 67(7):639, July 1999.

\bibitem{SchulmanBook}
L.~S. Schulman.
\newblock {\em Techniques and Applications of Path Integration}.
\newblock Dover Publications, Inc., Mineola, New York, 2005.
\newblock See Eq. (32.39) at page 327.

\bibitem{HausandPan}
H.~A. Haus and J.~L. Pan.
\newblock Photon spin and the paraxial wave equation.
\newblock {\em Am. J. Phys.}, 61(9):818, September 1993.

\bibitem{BickelandBailey}
William~S. Bickel and Wilburn Bailey.
\newblock Stokes vectors, mueller matrices, and polarized scattered light.
\newblock {\em Am. J. Phys.}, 53(5):468, May 1985.

\end{thebibliography}
%
\end{document}